\documentclass[12pt,preprint]{aastex62} 
\usepackage{color}
\usepackage{hhline}
\usepackage{graphicx}
\usepackage{booktabs}
\usepackage{threeparttable}
\usepackage{longtable}
\usepackage{rotating}
\usepackage{hyperref}
\usepackage{scalefnt}
\hypersetup{
    colorlinks=true,
    linkcolor=blue,
    filecolor=magenta,      
    urlcolor=blue,
}
\shorttitle{Fluorine Abundances in the Galactic Disk}
\shortauthors{Guer\c{c}o et al.}

\begin{document}

\title{Fluorine Abundances in the Galactic Disk}

\author[0000-0002-0151-5212]{Rafael Guer\c{c}o}
\affil{Observat\'orio Nacional, S\~ao Crist\'ov\~ao, Rio de Janeiro, Brazil}
\email{rguerco@on.br}

\author{Katia Cunha}
\affil{University of Arizona, Tucson, AZ 85719, USA}
\affil{Observat\'orio Nacional, S\~ao Crist\'ov\~ao, Rio de Janeiro, Brazil}

\author[0000-0002-0134-2024]{Verne V. Smith}
\affil{National Optical Astronomy Observatories, Tucson, AZ 85719 USA}

\author[0000-0003-2969-2445]{Christian R. Hayes}
\affil{University of Virginia, Charlotesville, VA 22904, USA}

\author[0000-0002-5665-2716]{Carlos Abia}
\affil{Depto. Fisica Teorica y del Cosmos, Universidad de Granada, E-18071 Granada, Spain}

\author[0000-0003-1814-3379]{David L. Lambert}
\affil{W. J. McDonald Observatory and Department of Astronomy, University of Texas at Austin, Austin, TX 78712, USA}

\author[0000-0002-4912-8609]{Henrik J\"onsson}
\affil{Lund Observatory, Department of Astronomy and Theoretical Physics, Lund University, Box 43, SE-22100 Lund, Sweden}
\affil{Materials Science and Applied Mathematics, Malm\"o University, SE-205 06 Malm\"o, Sweden}

\author[0000-0001-6294-3790]{Nils Ryde}
\affil{Lund Observatory, Department of Astronomy and Theoretical Physics, Lund University, Box 43, SE-22100 Lund, Sweden}

\begin{abstract}
 
The chemical evolution of fluorine is investigated in a sample of Milky Way red giant stars that span a significant range in metallicity from [Fe/H]$\sim$ -1.3 to 0.0 dex. 
Fluorine abundances are derived from vibration-rotation lines of HF in high-resolution infrared spectra near $\lambda$2.335$\mu$m. The red giants are members of the thin and thick disk / halo, with two stars being likely members of the outer disk Monoceros overdensity.  
At lower metallicities, with [Fe/H] $<$ -0.4 to -0.5, the abundance of F varies as a primary element with respect to the Fe abundance, with a constant subsolar value of [F/Fe]$\sim$-0.3 to -0.4 dex.  At larger metallicities, however, [F/Fe] increases rapidly with [Fe/H] and displays a near-secondary behavior with respect to Fe.  
Comparisons with various models of chemical evolution suggest that in the low-metallicity regime (dominated here by thick disk stars), a primary evolution of $^{19}$F with Fe, with a subsolar [F/Fe] value that roughly matches the observed plateau can be reproduced by a model incorporating neutrino nucleosynthesis in the aftermath of the core collapse in supernovae of type II (SN II). A primary behavior for [F/Fe] at low metallicity is also observed for a model including rapid rotating low-metallicity massive stars but this overproduces [F/Fe] at low metallicity.
The thick disk red giants in our sample span a large range of galactocentric distance (R$_{\rm g}$ $\sim$ 6--13.7 kpc), yet display a $\sim$constant value of [F/Fe], indicating a very flat gradient  (with a slope of 0.02 $\pm$ 0.03 dex/kpc) of this elemental ratio over a significant portion of the Galaxy having $|Z|>$ 300 pc away from the Galaxy mid-plane.

\end{abstract}


\section{Introduction}

The main processes responsible for the synthesis of chemical elements in the Universe include nucleosynthesis during the Big Bang and spallation reactions driven by cosmic ray interactions with the ambient interstellar medium, which produce much of the light-element abundances (H, He, Li, Be, B), while nuclear reactions in stellar interiors produce the bulk of the heavy elements, from carbon on up. Understanding the astrophysical origins of the different chemical elements provides insights into both the formation and evolution of stars, as well as the Galaxy and its various components, such as the disk, the bulge, or the halo. 
Fluorine is the focus of the work presented here, as its detailed origins remain somewhat uncertain, given that the behavior of the fluorine abundance as a function of metallicity (e.g., [Fe/H]), in particular at low metallicity, is not well known.

Fluorine consists of a single stable isotope, $^{19}$F, and has been suggested to be produced in a variety of astrophysical sites, such as asymptotic giant branch (AGB) stars via neutron and proton captures during He-burning thermal pulses (e.g, Jorissen et al. 1992; Forestini et al. 1992; Cristallo et al. 2014), inelastic neutrino scattering off of (primarily) $^{20}$Ne during Type II supernovae (SNII; Woosley \& Haxton 1988; Woosley et al. 1990), core He-burning in Wolf-Rayet stars undergoing high mass-loss rates (e.g., Meynet \& Arnould 2000), or rapidly rotating low-metallicity high-mass stars (e.g., Prantzos et al. 2018). 
In the AGB, Wolf-Rayet, and rapidly rotating low-metallicity high-mass stars, $^{19}$F is produced by a variety of $\alpha$-captures (He-burning), along with neutron and proton captures (requiring hydrogen to be mixed into the He-burning regions), but the expected reaction sequences are somewhat different between the low-mass AGB stars and the higher mass stars, although in all sources the reaction sequences leading to F production begin with $^{14}$N.  In the case of AGB stars, Cristallo et al. (2014) favor the reaction sequence $^{14}$N(n,p)$^{14}$C($\alpha$,$\gamma$)$^{18}$O(p,$\alpha$)$^{15}$N($\alpha$,$\gamma$)$^{19}$F as the major source of F, while for the higher mass stars, the expected main F production channel is $^{14}$N($\alpha$,$\gamma$)$^{18}$F($\beta^{+}$)$^{18}$O(p,$\alpha$)$^{15}$N($\alpha$,$\gamma$)$^{19}$F (Meynet \& Maeder 2000).
With such a variety of possible sources and somewhat varied nuclear reaction chains, the chemical evolution of fluorine in the Galaxy may be complex, but has been modelled by a number of authors, such as Renda et al. (2004), Kobayashi et al. (2011a; 2011b), Spitoni et al. (2018), and Prantzos et al. (2018). 

Several observational studies of F in red giants, including AGB stars, have demonstrated and documented the nucleosynthesis of $^{19}$F in AGB stars (Jorissen et al. 1992; Uttenthaler et al. 2008; Abia et al. 2009, 2010,
 2015, 2019), although the importance of AGB stars as a source driving the chemical evolution of F remains uncertain.  Fluorine has also been shown to be a reactant in the H-burning cycles that have altered the C, N, O, F, Na, Mg, and Al abundances between the different generations of stars found in globular clusters (Cunha et al. 2003; Smith et al. 2005; Yong et al. 2008; Alves-Brito et al. 2012; de Laverny \& Recio-Blanco 2013a; 2013b; D'Orazi et al. 2013), with the F abundances found to be lower in the so-called 2nd generation globular cluster stars.  
In the globular cluster stars, the above-mentioned abundance studies have generally found that the F abundances are found to anti-correlate with the Na abundances; these results may suggest that hot bottom burning (HBB) in massive AGB stars (M$\sim$4-6M$_{\odot}$) destroys $^{19}$F during Ne-Na H-burning at the base of the deep convective envelopes (e.g., Fenner et al. 2004; Smith et al. 2005).
The fact that AGB stars can be both a source of $^{19}$F during He-burning thermal pulses, as well as a sink in massive AGB stars during HBB illustrates one facet of the complexity in the chemical evolution of fluorine.

Attempts to probe the chemical evolution of fluorine in the Galaxy and to quantify the relative importance of the various astrophysical sources of $^{19}$F have relied on studies of field stars or open clusters across a range of metallicities.  Due to the lack of strong atomic fluorine lines in the optical or infrared (IR) parts of the spectrum, observations of HF are the only practical tool for deriving F abundances; however the HF lines are typically fairly weak, due to fluorine's intrinsically low abundance coupled with HF's rather modest dissociation energy (D$_{0}$=5.87eV). 
Due to these observational challenges, most of the studies directed at understanding the chemical evolution of fluorine have secured its behavior at roughly solar to moderately subsolar metallicities and have reached differing conclusions: Recio-Blanco et al. (2012), from an analysis of cool main-sequence stars with abundances from [Fe/H]$\sim$-0.4 to +0.1, conclude that neutrino nucleosynthesis (the $\nu$-process) in SN II appear to be the main site of F production.  Nault \& Pilachowski (2013) and Pilachowski \& Pace (2015), from a study of near-solar metallicity open clusters and field G and K giants also conclude that the $\nu$-process is an important source for Galactic F, with perhaps some contribution from AGB stars.  J\"onsson et al. (2017), using a sample of some 49 G and K giants having a metallicity range of [Fe/H]$\sim$-0.7 to +0.3, find that fluorine behaves as a secondary element relative to Fe and O and conclude that AGB stars are the dominant source of $^{19}$F in the Galaxy.
(The net increase in the abundance (the yield), dN(X), of secondary element X from its progenitor star is proportional to the initial metallicity (e.g., N(Fe)) of the star, so the final yield will be proportional to the metallicity squared (N(X)$\propto$N(Fe)$^{2}$).  The stellar yield of a primary element, Y, is independent of the progenitor star's metallicity and will increase linearly with the stellar metallicity (N(Y)$\propto$N(Fe))). 

Probing the behavior of fluorine over a larger range of metallicity, in particular pushing to low metallicities (i.e., below [Fe/H]$\sim$-0.7 dex) would help to elucidate the origins of fluorine: this is the main goal of this study.
This paper is organized as follows: the target selection and observations of the infrared and optical spectra are discussed in Section 2; Section 3 describes the determination of the stellar parameters and Section 4 the abundance analysis and uncertainties in the abundance determinations. The results are compared to literature values in Section 5 and discussed in Section 6.

\section{Target Selection and Observations}
 
The chemical evolution of fluorine in the Galaxy requires a sample of cool giants that includes stars with low metallicity, yet have detectable HF lines; this requires a search for red giants with both cool effective temperatures (T$_{\rm eff}$) and low values of [Fe/H].  Such combinations become rare in field giants due to the shift of low-metallicity giants at constant luminosity to higher values of T$_{\rm eff}$ relative to higher-metallicity giants.  The vibration-rotation HF lines in the K-band, near $\lambda$2.3$\mu$m, are the only spectral lines with which to determine precise fluorine abundances.  These HF lines can be detected in red giants with effective temperatures of $\sim$4000K or less and metallicities in the range of [Fe/H]$\sim$ -0.7 to -1.5, via high-resolution IR spectra (R$\sim$45,000) with signal-to-noise ratios $>$100. 
 
The target stars were selected as follows: thirteen nearby red-giants were selected from the FTS KPNO archive; four targets (observed with the Phoenix spectrograph on Gemini South) were selected from the GGSS Survey (Kundu et al. 2002); five targets (observed with the Phoenix spectrograph at the KPNO 2.1 m telescope) were selected from Norris et al. (1986); and five targets (observed with the iShell spectrograph on the IRTF) were selected from the APOGEE DR14 survey (Majewski et al. 2017).
The selected red giants in this study are found in Table \ref{tab:observations}; their parallaxes are from Gaia DR2 (Gaia Collaboration 2018) and distances from Bailer-Jones et al. (2018) are also listed (when available), along with their V and K magnitudes, which were used in the effective temperature determinations of some of the sample stars, plus the reddening A$_{\rm v}$, when considered. 
This sample contains red giants from the Galactic disk, probably from both the thin and thick disks; their Gaia parallaxes and distances indicate that they cover a large range of distances, going from the very local neighborhood, including the nearest red-giant Arcturus, to more distant stars (d $\sim$ 6 kpc). Their metallicities also encompass the metallicity range of the thin disk, with overlap of the thick disk, overall spanning a range from roughly solar ([Fe/H]=0) to [Fe/H]=-1.2 dex. 
 
\subsection{Infrared Spectra}

No useful atomic fluorine lines are available in the visible region of the spectrum and detectable atomic F I lines lie in the far UV at $\lambda\sim$ 950 \AA. However, in cool stars, fluorine abundances can be derived using the HF molecule, which has transitions in the infrared that provide detectable lines from its fundamental vibrational mode. 

In order to study the HF rotational-vibrational transitions near $\lambda$2.358 $\mu$m in the spectra of red giants, high-resolution spectra were obtained with combinations of spectrographs and telescopes; some instrumental set-ups allowed only for the observation of the HF (1-0) R9 line, while in others secondary HF lines could be observed.
The observed red giants and corresponding spectrograph/telescope combinations used are listed in Table \ref{tab:observations}.

Observations with the NOAO Phoenix Spectrograph (Hinkle et al. 2003) were obtained on two different telescopes: the Kitt Peak National Observatory (KPNO) 2.1-meter and the Gemini South 8.1-meter Telescope. In addition, the 3.0-meter NASA Infrared Telescope Facility (IRTF) was used with the iSHELL spectrograph (Rayner et al. 2016). 
The spectra obtained with iSHELL on the IRTF were geared towards observations of low metallicity stars based on SDSS/APOGEE (Majewski et al. 2017) DR14 stellar parameters, while the Phoenix observations were aimed primarily at red giants having intermediate metallicities, as determined from Washington photometry in the Grid Giant Star Survey (GGSS; Kundu et al. 2002).

The Phoenix spectra were centered at $\sim\lambda$23,360\AA \ and have a resolution R $\equiv$ $\lambda/\Delta\lambda$ = 50,000. The Phoenix spectra have a limited spectral coverage of $\sim$100\AA; although small, the selected region contains molecular lines of CO, the HF(1-0) R9 transition and, in particular for the Gemini-S spectra, it contains an isolated atomic Fe I line at 23,308.477\AA \ that can (in the absence of spectra with larger spectral coverage) be used to estimate the stellar metallicity. 
It should be noted, however, that the Phoenix spectra obtained with the KPNO 2.1-m have a slightly different wavelength coverage (from $\lambda$23,315 to 23,430 \AA) and do not contain the Fe I line. 
The reduction of all the obtained Phoenix spectra to one dimension followed standard procedures; see the discussion in Cunha et al. (2003). 

The iSHELL spectra analyzed here have higher resolution than the Phoenix spectra (R $\equiv$ $\lambda/\Delta\lambda$ = 75,000) and much larger spectral coverage, between $\sim$ 22,850 and 23,750 \AA. 
The available Spextool v5.0.2 (SPectral EXtraction TOOL; Cushing, Vacca \& Rayner 2004) was used to perform the reduction of the iSHELL spectra to one dimension. The reduction steps include: creation of normalized flat field images and wavelength calibration files; a non-linearity correction was applied to the raw data and flat fields; apertures positions 
 were identified and extracted; the raw data were divided by the flat fields; 
telluric corrections and flux calibrations were performed on combined multi-order spectra; telluric-corrected spectra were merged into a single continuous spectrum for each star; finally the continuous spectra were corrected by removing bad pixels.

In addition to the Phoenix and iSHELL spectra, we collected IR spectra from the Fourier Transform Spectrometer (FTS) archive (Pilachowski et al. 2017;  \url{https://sparc.sca.iu.edu}), 
for a sample of disk red giants observed with the FTS on the KPNO 4-m telescope (Hall el al. 1979).
The FTS spectra analyzed have R $\equiv$ $\lambda$/$\Delta\lambda$ = 45,000 and cover the spectral region between $\lambda$19,500--24,000\AA. 
Most of the selected FTS targets are bright, nearby red giants analyzed previously for chemical abundances by Smith \& Lambert (1985; 1986; 1990) and Jorissen et al. (1992).
We also analyzed the higher resolution (R = 100,000) FTS spectrum  obtained with the 4-m telescope of the ``standard'' red giant Arcturus (Hinkle, Wallace \& Livingston 1995; \url{ftp://ftp.noao.edu/catalogs/arcturusatlas/ir/}).
All spectral data sets included observations of hot standard stars observed on the same nights (except for two stars with observations from the FTS archive) whenever possible over a range of air masses. These were used to map the telluric lines in this spectral region (consisting of a mixture of CH$_{4}$ and H$_{2}$O lines), which were then removed by dividing the program stars by the hot star, taking into account airmass differences.  The task `telluric' within the software package IRAF was used, except for the iSHELL spectra, where telluric division was included as part of the reduction pipeline.  In addition, all stellar and telluric spectra were inspected visually in the regions of the HF lines to ascertain the significance of telluric contamination to each line.

\begin{figure}[!htb]
   \centering
   \includegraphics[width=1.0\textwidth]{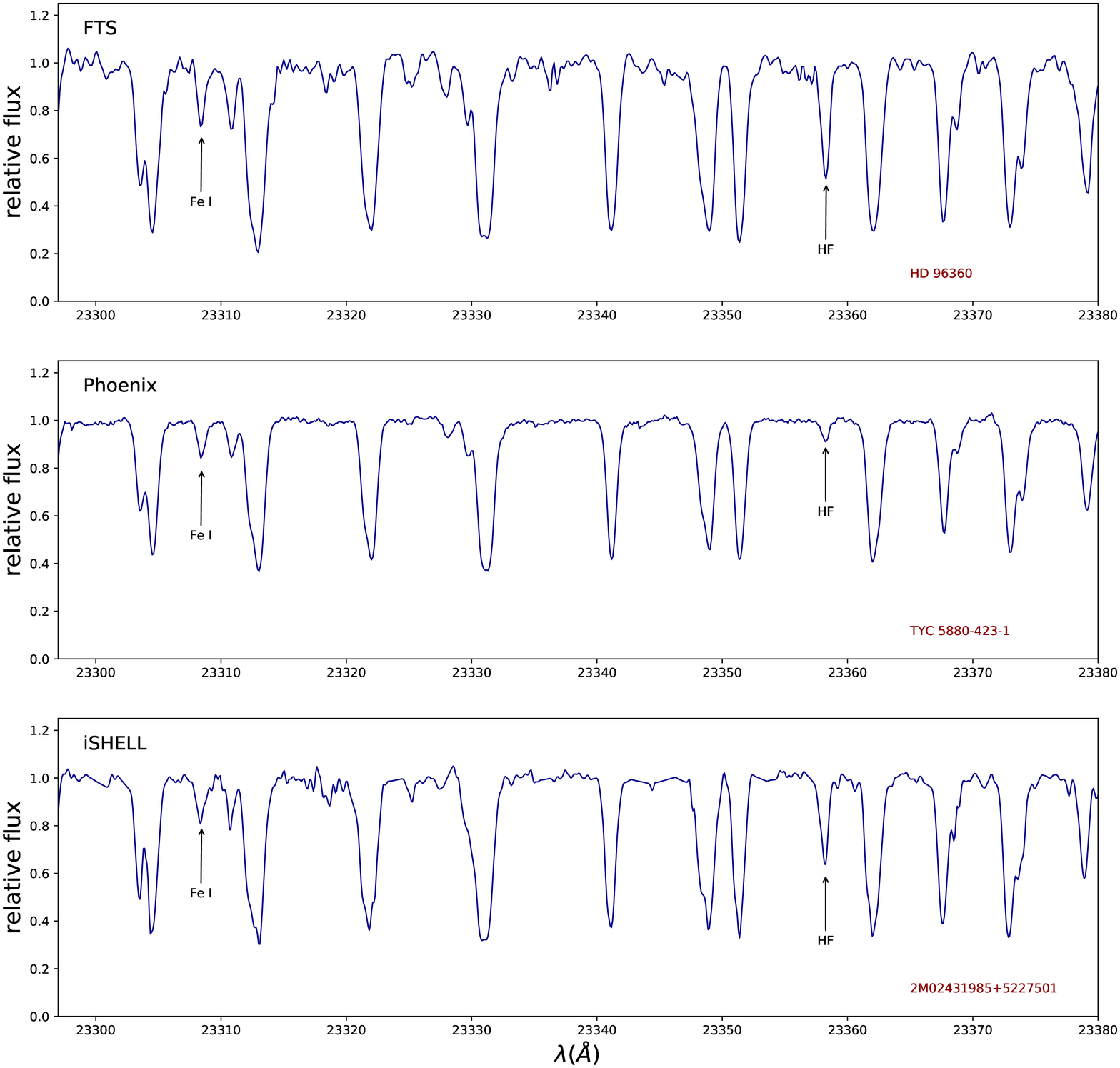}
   \caption{Sample spectra for each one of the spectrographs used in the observations: the FTS on the KPNO 4-m (top panel), Phoenix on Gemini-S (middle panel), and iSHELL on the IRTF (bottom panel). The selected spectral region includes the HF(1-0) R9 line at 2.3$\mu$m and one of the Fe I lines analyzed, both identified in the spectra.}
   \label{fig:observed}
\end{figure}

Figure \ref{fig:observed} shows sample reduced spectra for each one of the three different infrared spectrographs used in the observations: FTS (top panel), Phoenix (middle panel), and IRTF (bottom panel). The spectra are shown in the region between $\sim$ $\lambda$23,300--23,400\AA, most of which is available for all targets.

\subsection{Optical spectra}

Four of our target stars (HD 19697, HD 20305, HD 28085, and HD 90862) were also observed with the high-resolution optical Sandiford Cassegrain Echelle Spectrometer (SES) on the 2.1-m Struve Telescope at McDonald Observatory; the SES spectra analyzed have a resolution R = 60,000 and cover wavelength range between $\lambda$6030--7980\AA. The optical spectra of these four targets were used to derive the metallicties for the stars given that their Phoenix spectra (obtained on the KPNO 2.1-m telescope) had limited wavelength coverage, not reaching the Fe I line at $\lambda$23,308.477\AA.
The reduction of the SES spectra to one dimension (described in Bizyaev et al. 2006) used the IRAF package and standard steps for bias subtraction, division by flat field, scattered light subtraction, as well as, wavelength calibration using a Th-Ar lamp.

The optical spectrum of the ``standard'' red giant Arcturus (Hinkle et al. 2000) was also analyzed here with the goal of comparing the metallicity scales obtained from the analysis of the optical and infrared spectra and evaluate possible systematic offsets between the derived metallicities. The results from this comparison will be presented in Section 5.

\section{Stellar Parameters} \label{sec:parameters}

\begin{figure}[t!]
  \centering
  \includegraphics[width=1.0\textwidth]{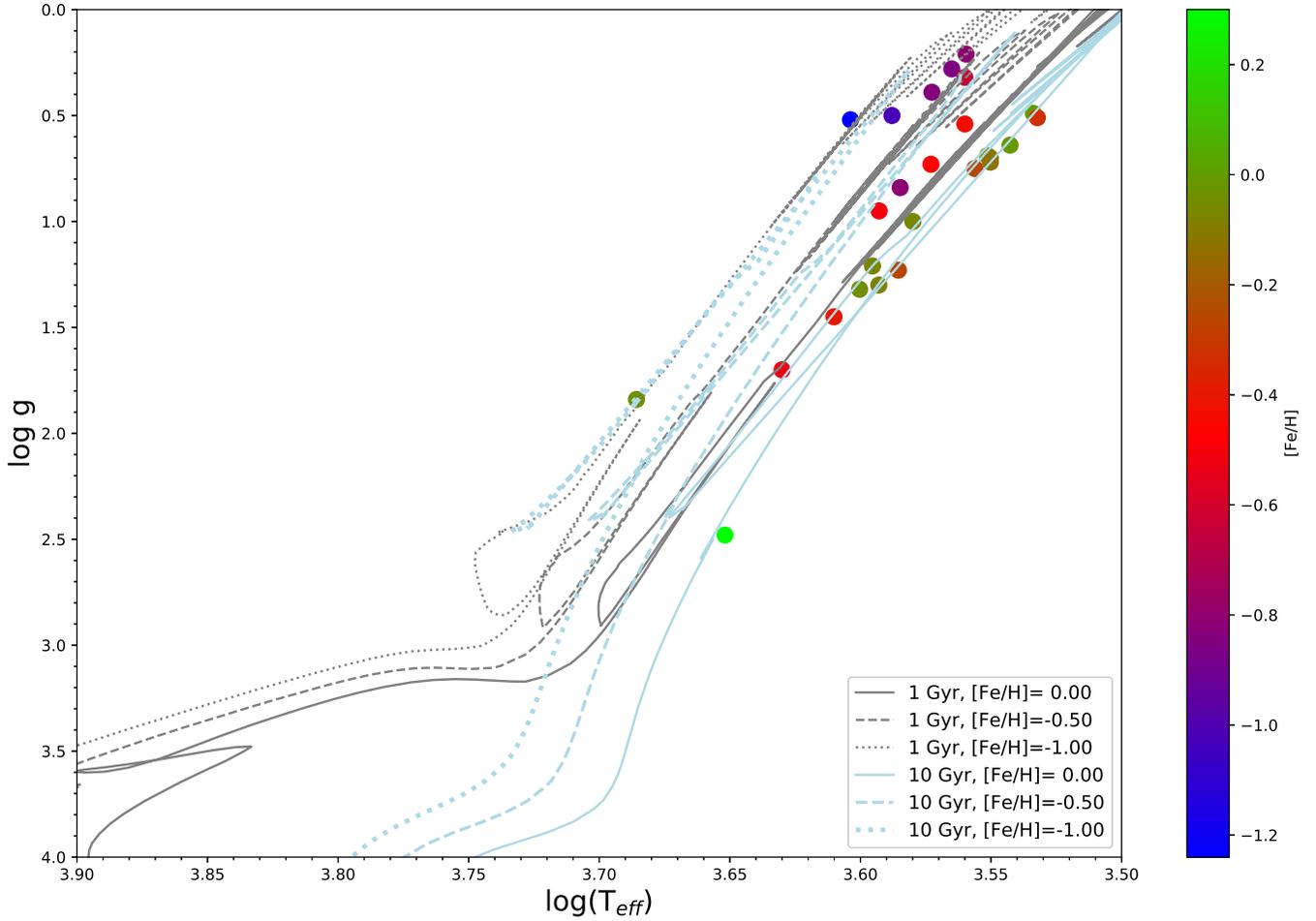}
  \caption{The T$_{\rm eff}$ - Log g diagram (Kiel Diagram) for the red giants in this sample; isochrones from Bressan et al. (2012) for solar (solid lines); -0.5 (dashed lines) and -1.0 (dotted lines) metallicities; ages 1 Gyr (gray lines) and 10 Gyr (blue lines) are also shown as a context to illustrate relative positions on the red giant branch. The stars are color coded by metallicity.}
  \label{fig:HR}
\end{figure}

A T$\rm _{eff}$ - log g diagram showing the stars studied here is presented in Figure \ref{fig:HR}; the target red giant stars are quite luminous and located high up on the red-giant branch. For reference we also show in the figure the PARSEC (PAdova and TRieste Stellar Evolution Code) isochrones from Bressan et al. (2012) with metallicities corresponding roughly to the range in metallicity of the studied stars: solar, -0.5 and -1.0. 
The stellar parameters derived for the studied stars are presented in Table \ref{tab:parameters}. In the following section we discuss the stellar parameters determinations.

\subsection{Effective Temperatures}

The effective temperatures for the studied stars were obtained either from a photometric calibration, from optical lines of Fe I, or from the H-band APOGEE spectra, depending on the availability of spectra, reliability of photometric magnitudes, and degree of extinction. 

For those target stars not having optical nor APOGEE spectra, 
we derived photometric T$\rm _{eff}$s from the V--K colors (Table \ref{tab:observations}), using the calibration in Bessell et al. (1998; T$\rm _{eff}$ = 9102.523 - 3030.654(V--K) + 633.3732(V--K)$^2$ - 60.73879(V--K)$^3$ + 2.135847(V--K)$^4$). For two stars in the sample (2MASS06232713-3342412 and 2MASS07313775-2818395) V magnitudes were not available and we used the 2MASS J-K magnitudes, transformations from Carpenter (2001) and the same calibration (Bessell et al. 1998) to estimate the effective temperatures.
For the nearby stars in our sample interstellar extinction was not considered; for stars with distances larger than 150 pc and less than 1 kpc, we determined A$_V$ using the Chen et al. (1998a) extinction law with the reddening correction E(B-V) and extinction A$_K$ determined using R$_V$ = 3.070 and R$_K$ = 0.342 from McCall (2004).
For more distant stars (having distances larger than $\sim$ 1 kpc, Table \ref{tab:observations}, we relied on Green et al. (2018), who provide E(B-V) using 3D Dust Mapping with Pan-STARRS 1 and 2MASS data. 

For the target stars taken from the SDSS IV - APOGEE survey (observed with the iSHELL spectrograph), we adopted the APOGEE DR14 calibrated effective temperatures derived from the APOGEE spectra automatically by the ASPCAP pipeline (Garc\'ia P\'erez et al. 2016). ASPCAP derives stellar parameters and metallicities from synthetic spectra and global 7-D fits to the APOGEE spectra via $\chi$-squared minimization.

The effective temperatures for those stars having SES spectra (HD 19697, HD 20305, HD 28085 and HD 90862; Table \ref{tab:observations}) were determined using a sample of Fe I lines between 6050 \AA\ and 6860 \AA\ and their measured equivalent widths. The methodology adopted in the determination of the stellar parameters derives at the same time the stellar metallicities and will be discussed in Section 4.2.

\subsection{Surface Gravities}

The surface gravities for the stars were derived using the PARAM 1.3 code (\url{http://stev.oapd.inaf.it/cgi-bin/param_1.3}) with parallaxes, V magnitudes, effective temperatures, initial values for the metallicities as inputs (Table \ref{tab:observations}), and using the PARSEC isochrones from Bressan et al. (2012) for scaled-solar composition, with the Y=0.2485+1.78Z relation and Z$_\odot$=0.0152.
The PARAM 1.3 code is a web interface for the Bayesian estimation of stellar parameters (see Rodrigues et al. 2014 and da Silva et al. 2006).
The parallaxes for the stars were taken from Gaia DR2.
In a few cases, however, parallaxes were not available in Gaia - DR2 and Hipparcos parallaxes (van Leeuwen 2007) were used instead (distances for those stars with parallaxes only in Hipparcos were determined by simply the inverse of the parallax). 

For those targets with results available in APOGEE DR14, we can compare our derived log g's obtained using the PARAM 1.3 code, with those obtained by the APOGEE team using APOGEE spectra and the APOGEE stellar parameters abundance pipeline (ASPCAP; Garc\'ia P\'erez et al. 2016). It should be noted, however, that the ASPCAP surface gravities for red giants have systematic offsets. These are, however, well-calibrated using log values from seismic studies; the APOGEE team also releases calibrated values of log g (see the discussion in Holtzman et al. 2018).
The mean difference between the log g values adopted in this study (from PARAM 1.3) with the calibrated log g values from APOGEE DR14 is $<$ $\delta$ log g$>$ = -0.01 $\pm$ 0.11.

\section{Abundance Analysis} \label{sec:abundance}

The stellar atmospheric models used in all of the abundance calculations presented in this study are spherical models from the MARCS grid (Gustafsson  et  al. 2008). 
Given the stellar parameters for the stars we interpolated models using the online grid of OSMARCS models (\url{http://marcs.astro.uu.se/}) for M = 1 M$_\odot$, $\xi$ = 2.0 km$\cdot$s$^{-1}$, and standard chemical composition (standard composition in this case corresponds to solar composition models for solar metallicity stars and $\alpha$-enhanced models for low metallicity stars).  
The microturbulent velocities for the target stars were estimated
from a relation derived in Guer\c{c}o et al. (2019), that was obtained using the $\xi$ and log g results from the study by Alves-Brito et al. (2010). (Guer\c{c}o et al. 2019 also tested the microturbulent relation proposed in Pilachowski et al. 1996 and found that the microtubulent velocity results agreed well.)

\subsection{Analysis of Infrared Spectra}

Synthetic spectra were calculated in Local Thermal Equilibrium (LTE) with the code Turbospectrum (TS; Alvarez \& Plez 1998; Plez 2012), noting that TS is compatible with the adopted spherical model atmospheres given its spherical radiative transfer treatment.  
Best fits between synthetic and observed spectra were obtained and the synthetic spectra were manipulated using the plotting and broadening routines from the MOOG code (Sneden 1973).
The model spectra were broadened by convolution with a Gaussian profile, which represents the combination of the instrumental profile, macroturbulent velocity, and projected rotational velocity (v$\cdot$sin$i$; it is expected that the v$\cdot$sin$i$ values for red giants should be small).

\subsubsection{Fluorine Abundances}

\begin{figure}[t!]
  \centering
  \includegraphics[width=1.0\textwidth]{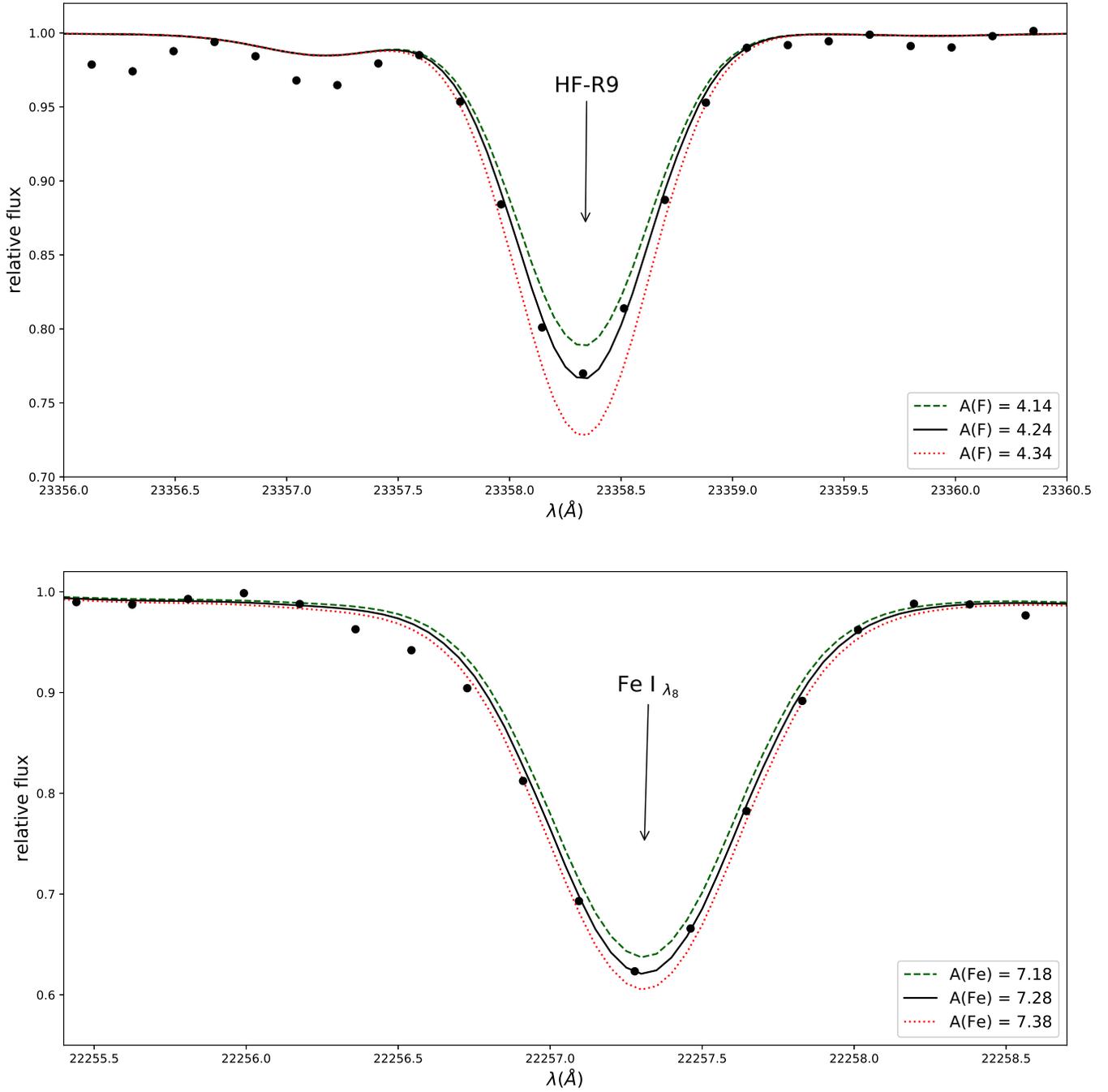}
  \caption{Top Panel: The observed and synthetic spectra of the star 41 Com showing the HF(1-0) R9 transition. The syntheses were computed for: A(F) = 4.14 (green dashed line), 4.24 (best synthesis; solid line) and 4.34 (red dashed line). Bottom Panel: same for the atomic Fe I line at 22,257.107\AA\ for A(Fe) = 7.18; 7.28 and 7.38 dex.}
  \label{fig:synthesis}
\end{figure}

Fluorine abundances were derived from a sample of up to five vibrational-rotational HF (1-0) transitions: R9, R13, R14, R15 and R16, depending on the spectral coverage for each target star.
Table \ref{tab:lines} presents the HF lines used in the abundance analyses, along with the excitation potential ($\chi$; from J\"onsson et al. 2014a and Decin 2000), log $gf$ (from J\"onsson et al. 2014a), and dissociation energy (D$_\circ$=5.869 eV; from Sauval \& Tatum 1984) of the transitions.  We note that there remains some uncertainty in the value of D$_\circ$ for HF, as Luo (2007) has reported a value of D$_\circ$=573.398$\pm$0.011 kJ/mol, or 5.943 eV/molecule.  This difference of 0.074 eV would lead to a slightly lower absolute F abundance of $\sim$0.04 dex for a typical red giant from this sample.  
To our knowledge no non-LTE calculations have been made for the HF transitions for stellar atmosphere conditions. However,  vibrational-rotational transitions in molecules in the ground electronic state should, in general, be in LTE (Hinkle \& Lambert 1975).
Sample synthetic profiles and best fit syntheses for the studied HF lines are presented in Figure \ref{fig:synthesis}.


\subsubsection{Iron Abundances}

Iron abundances were determined using a sample of 19 Fe I lines in the wavelength range between $\sim\lambda$19,900 -- 23,700\AA, with Fe I lines in the full wavelength range used for the stars observed with iSHELL.
Table \ref{tab:lines} presents the Fe I line list selected for the abundance analysis and the relevant atomic data. 
The log $gf$-values of these Fe I transitions were determined astrophysically, using the solar spectrum as a reference. 
The solar spectrum analyzed was the very high-resolution (R = 40,000) center-of-disk intensity spectrum obtained from: \url{http://www.eso.org/sci/facilities/paranal/decommissioned/isaac/tools/spectroscopic_standards.html}.
A canonical solar MARCS model atmospheres with T$\rm _{eff}$ = 5770 K; log g = 4.44, and $\xi$= 1.0 km/s was adopted in the calculations, and solar $gf$-values were obtained assuming a solar abundance of A(Fe) = 7.45 (Asplund et al. 2005). 

\begin{figure}[!htb]
   \centering
   \includegraphics[width=1.0\textwidth]{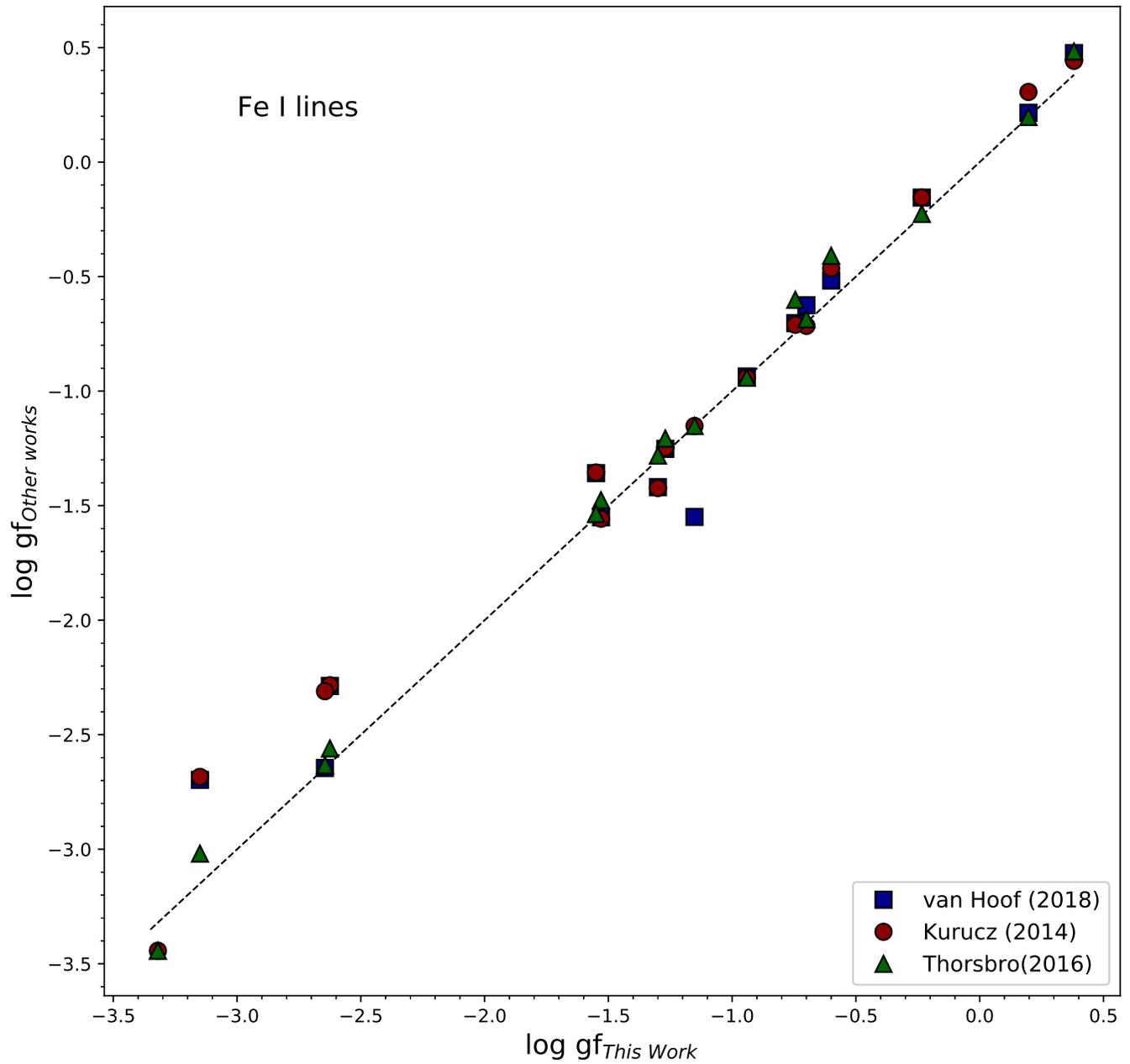}
   \caption{The comparison between the solar $gf$-values for the Fe I lines derived in this work with solar $gf$-values from Thorsbro (2016), as well as $gf$-values from the Kurucz (2014) line list and from van Hoof (2018).}
   \label{fig:loggf}
\end{figure}

In addition to the astrophysical $gf$-values, the van der Waals damping constants, $\Gamma_6$, were adjusted using both the TS code (consistent with the abundance analysis in this study), as well as with the MOOG code. The reason behind using two codes for the calculations of LTE synthetic spectra was to evaluate possible differences in the solar syntheses, given that the current version of TS does not consider Stark broadening in the computation of synthetic spectra. The $gf$-values derived using the two codes were similar, with mean differences in the log $gf$-values of $<\Delta$(log gf)$>$ = -0.02 $\pm$ 0.07. 
Most of the studied Fe I transitions have log $gf$ values larger than -1.5 dex. As shown in Figure \ref{fig:loggf}, our solar log $gf$ values are in good agreement with another set of solar $gf$-values obtained independently by Throsbro (2016) using the LTE synthesis code Spectroscopy Made Easy (SME; Valenti \& Piskunov 1996): the mean difference ``This Work - Throsbro'' $<\delta$(log $gf$)$>$ = -0.05$\pm$0.06; rms = 0.08.
The agreement with the values found in the Kurucz (1994) line list is also good, in particular for the stronger lines, while some of the weaker lines show larger discrepancies.

The individual Fe I line abundances obtained in this study are presented in Table \ref{tab:abundance}. Given the non-homogeneous set of observed spectra in this study, the number of Fe I lines analyzed to obtain the stellar metallicities was varied and depended on the spectral coverage available in each case. For a few stars only one Fe I line 23308 \AA \ could be measured and we verified that the Fe abundances from this Fe I line in particular were not systematically offset when compared to the abundances from the other Fe I lines.

\subsection{Analysis of Optical Spectra}
\subsubsection{Iron Abundances}

For the four stars in this sample with infrared spectra not having spectral coverage that includes any Fe I lines, their iron abundances were derived from measured equivalent widths of optical Fe I lines using the same LTE code TS.
Fe I lines in the spectral region covered by the SES spectra were selected from the master line list presented in the recent study by Ghezzi et al. (2018); the $gf$-values in that study are solar $gf$-values obtained for an assumed solar reference abundance A(Fe)$_{Sun}$= 7.50 (Asplund et al. 2009).
The selected line list containing 53 Fe I lines and the atomic data adopted are presented in Table \ref{tab:lines2}. 

\begin{figure}[!htb]
   \centering
   \includegraphics[width=1.0\textwidth]{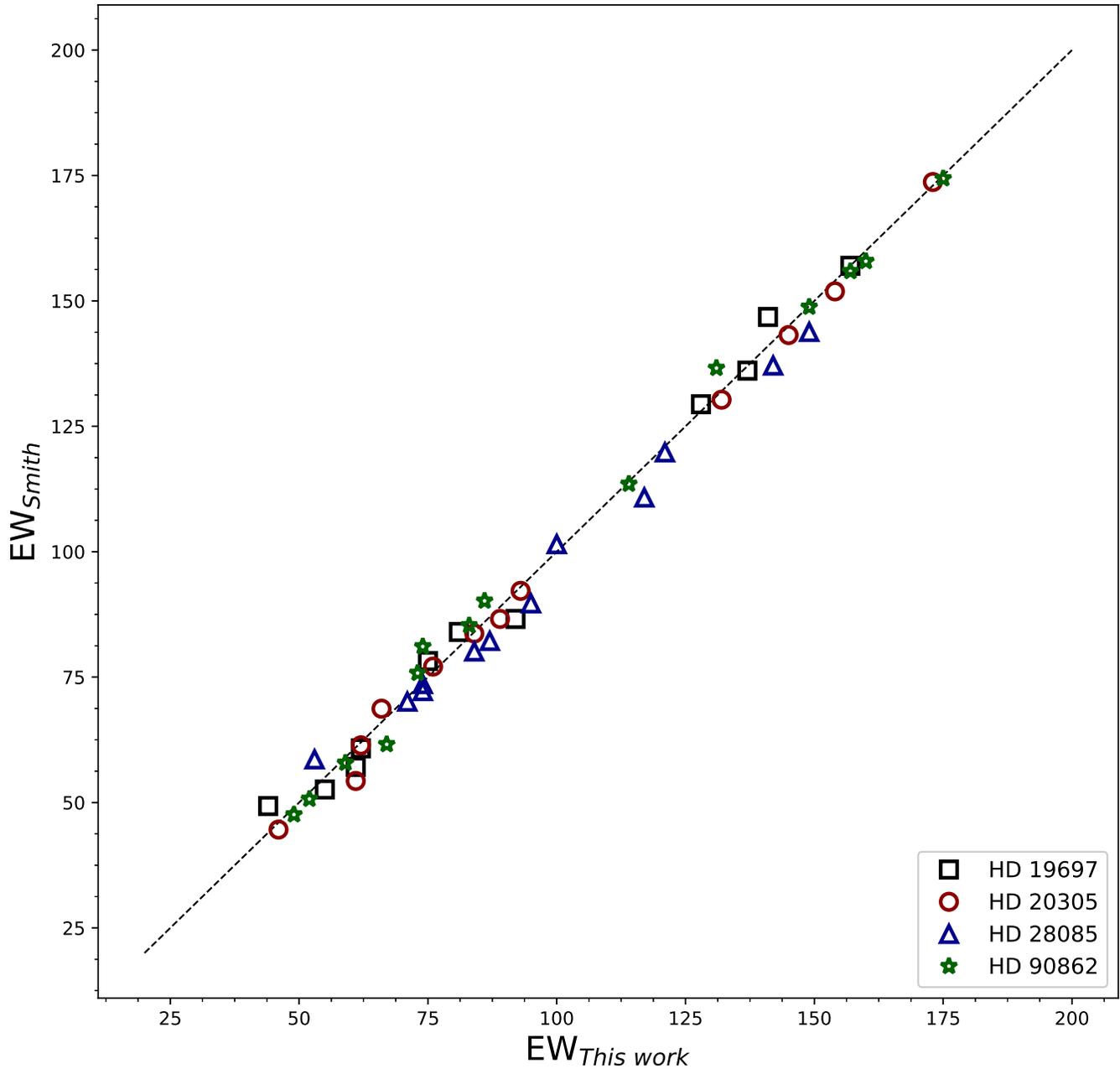}
   \caption{A comparison showing that there is good agreement between the equivalent widths of optical Fe I lines measured in this work with those measured previously by co-author V. Smith. The expected uncertainties in the measurements, given the spectral resolution and the signal-to-noise ratio (Cayrel 1988) is $\sim\pm$2 m\AA.}
   \label{fig:EW}
\end{figure}

The equivalent widths for the Fe I lines were measured with the IRAF splot package (Tody 1993) and assuming Gaussian profiles. 
Equivalent widths for a subset of the sample Fe I lines had been previously measured by V. Smith. 
A comparison of the equivalent width measured by these two co-authors is shown in Figure \ref{fig:EW}; the agreement is good without obvious offsets. The average of the differences between the equivalent width measurements in this study minus those measured by V. Smith is 0.6$\pm$3 m\AA \ with rms = 3.4 m\AA. 

\begin{figure}[!htb]
   \centering
   \includegraphics[width=1.0\textwidth]{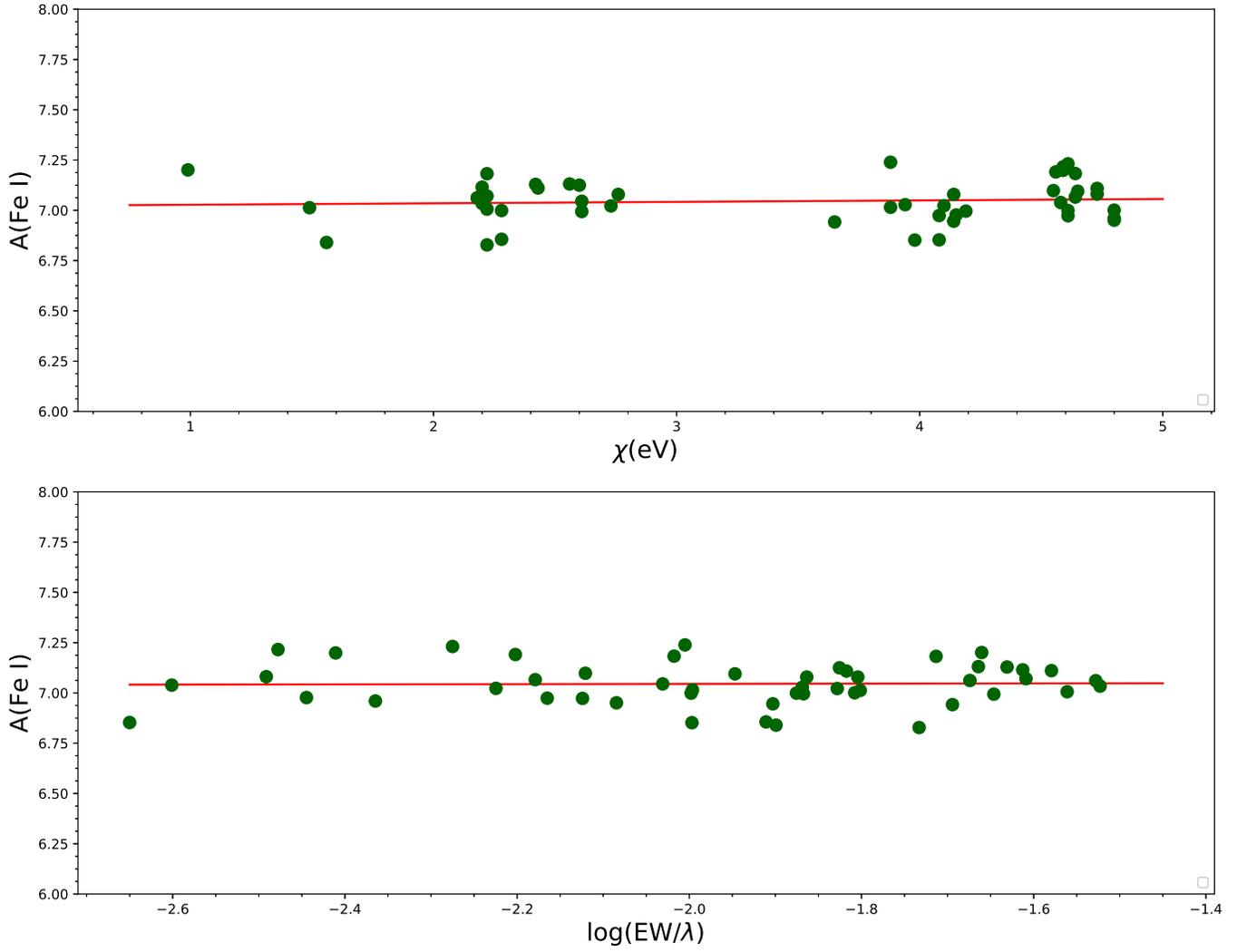}
   \caption{Top Panel: the run of the Fe abundances versus the excitation potential, $\chi$, of the Fe I lines in HD 20305; a zero slope defines the effective temperature for the star. Bottom Panel: iron abundances versus the reduced equivalent widths (log(EW/$\lambda$) for the Fe I lines; a zero slope here defines the microturbulent velocity ($\xi$).}
   \label{fig:parameters}
\end{figure}

The adopted methodology for analyzing the optical spectra derives, simultaneously, the effective temperature, the mean iron abundance, and the microturbulent velocity for each star. 
The effective temperature is defined when reaching an absence of trend (zero slope) in the diagram of the excitation potential ($\chi$) of the Fe I lines versus their individual iron abundances, while the microturbulent velocity parameter is defined from the requirement of finding zero slope in the equivalent width versus iron abundance diagram.
The log g values were derived with the code PARAM 1.3.
Each calculation was started with an estimated log $g$ value for the star and iterated until the results for the effective temperature, metallicity, microturbulent velocity, and log $g$ (obtained with the code PARAM 1.3; Section 3.2) were all consistent. 
Figure  \ref{fig:parameters} shows an example of the solution obtained for the star HD 20305: the top panel of the figure displays the effective temperature diagram and the bottom panel the microturbulent velocity diagram.

\subsection{Abundance Uncertainties}

Uncertainties in the derived abundances can be estimated from the sensitivities of the Fe I and HF measured abundances to changes in the stellar parameters corresponding to typical errors in T$\rm_{eff}$ ($\pm$100 K), log $g$ ($\pm$0.25 dex), metallicity ($\pm$0.1 dex), and microturbulent velocity ($\pm$0.3 km$\cdot$s$^{-1}$). 
Table \ref{tab:disturbance} has the estimated uncertainties, $\Delta$A (the square root of the sum in quadrature of the corresponding abundance changes $\delta$T$\rm_{eff}$, $\delta$log g, $\delta$[Fe/H] and $\delta \xi$), for baseline model atmospheres representative of the studied sample: one more metal-rich (T$\rm_{eff}$ = 3917 K, log g = 1.30, [Fe/H]= -0.08 dex, $\xi$ = 1.75 km$\cdot$s$^{-1}$) and another more metal-poor (T$\rm_{eff}$ = 3725 K, log g = 0.30, [Fe/H]= -1.10 dex, $\xi$ = 1.70 km$\cdot$s$^{-1}$). 
We also present the sensitivity of the optical Fe I lines for a baseline model with T$\rm_{eff}$ = 4075 K, log g = 1.45, [Fe/H]= -0.40 dex, $\xi$ = 1.67 km$\cdot$s$^{-1}$.
As discussed in Guer\c{c}o et al. (2019), errors in the effective temperatures account for most of the uncertainties in the fluorine abundances. 
In addition, we estimate an uncertainty of ~0.05 dex (or in many cases less) for the abundance uncertainty due to uncertainties in placement of the continuum.
While the fluorine abundances are most sensitive to T$\rm_{eff}$, the derived iron abundances show more sensitivity to log g and $\xi$. 
The standard deviations of the mean derived fluorine and iron abundances (corresponding to line-to-line abundance scatter) are presented in Table \ref{tab:parameters}; these are typically less than 0.1 dex.


\section{Results}

\begin{figure}[t!]
  \centering
  \includegraphics[width=1.0\textwidth]{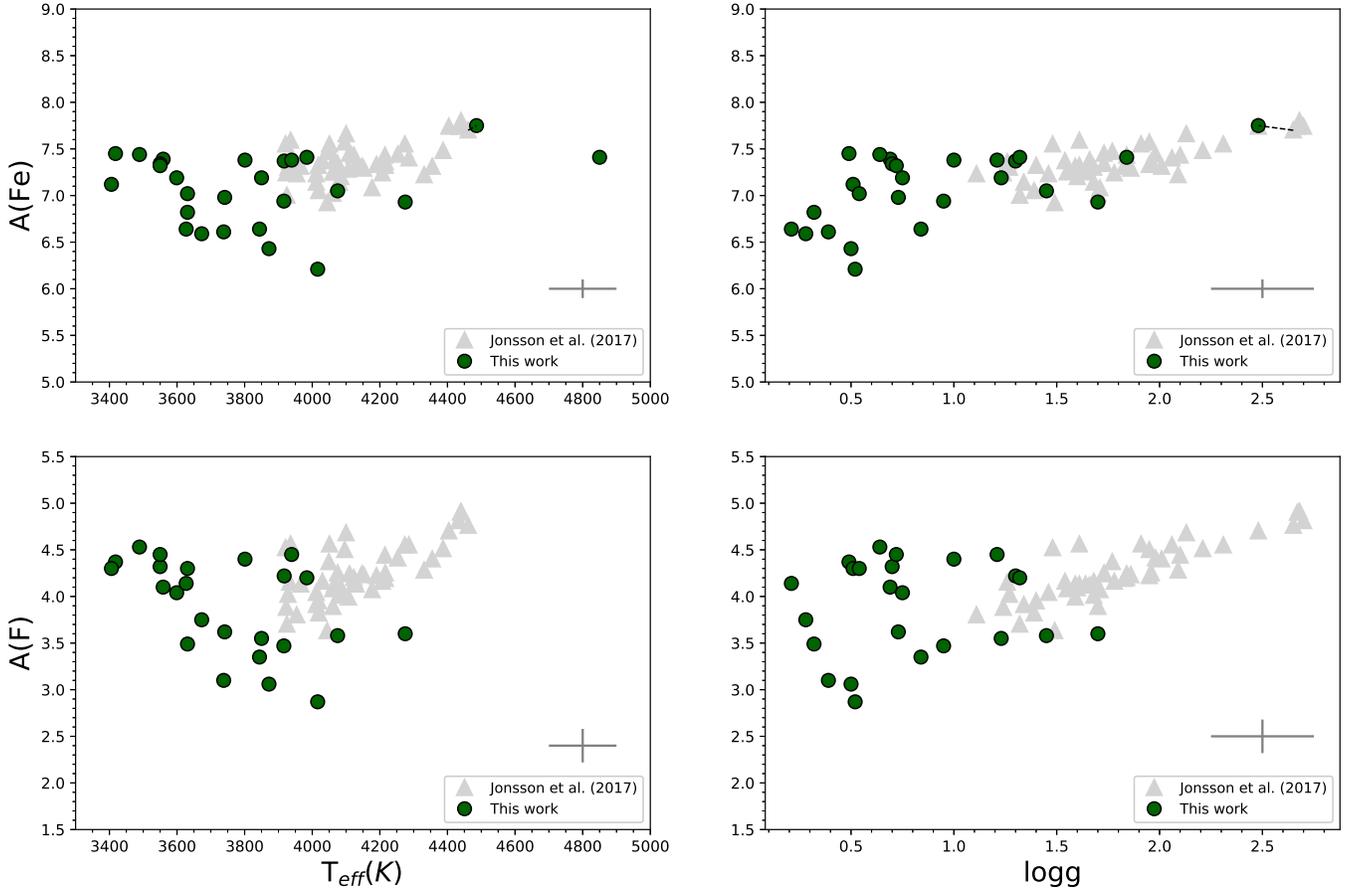}
  \caption{The abundances of F and Fe versus the derived stellar parameters of effective temperature (T$_{\rm eff}$ and surface gravity (log g) for this study and for J\"onsson et al. (2017). The Fe abundance results from each study for $\mu$ Leo are connected by the dashed line.  This figure illustrates the attempt in this study to include a larger fraction of cooler, lower surface gravity, and more metal-poor stars in which to detect HF in comparison to previous studies, using J\"onsson et al. (2017) as the example.}
  \label{fig:Teff-logg_vs-Fe-F}
\end{figure}

The mean F and Fe abundances and standard deviations of the individual line results are presented in Table \ref{tab:parameters}).
The F and Fe abundances as functions of the derived effective temperatures and log gs are shown in Figure \ref{fig:Teff-logg_vs-Fe-F}; the filled green circles represent the results from this study and for comparison we also show, as grey filled triangles, the fluorine abundances and metallicities from J\"onsson et al. (2017).
This comparison demonstrates that the range in stellar parameters for the stars in this study extends to much cooler effective temperatures and more luminous red giants than the sample from J\"onsson et al. (2017). In addition, this sample covers a larger range in metallicity, encompassing $\sim$1.2 dex, and an even larger range in fluorine. The abundance results do not show a significant trend in the iron abundances with T$_{\rm eff}$; an overall trend is observed with log g, which is expected, given that cool red giants with lower metallicities will have lower gravities and higher effective temperatures. The trend of the fluorine abundances with log g has larger scatter, showing two groups, one with lower F abundances and another with higher F abundances. 

\begin{figure}[t!]
  \centering
  \includegraphics[width=\textwidth]{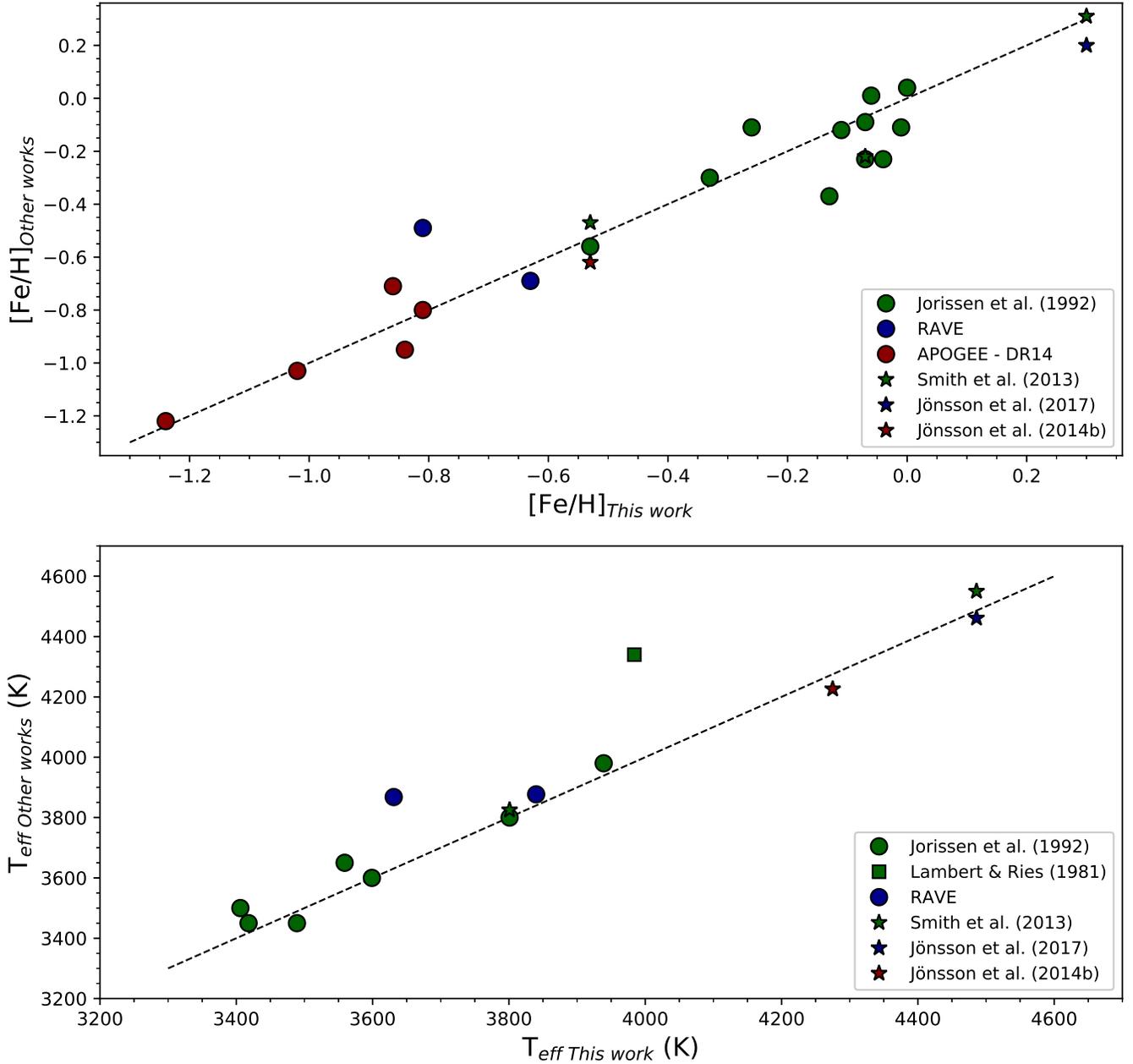}
  \caption{Top Panel: A comparison of the metallicities ([Fe/H]) derived in this study with results from the literature. The metallicities from RAVE are from Kordopatis et al. (2013) and from APOGEE are from DR14 (Holtzman et al. 2018). The expected uncertainties in the derived values of [Fe/H] are $\sim\pm$0.10 dex. There are no significant systematic differences between the iron abundances derived here and those from previous studies; there is only a small mean offset and standard deviation of +0.05$\pm$0.11 in [Fe/H] of this study minus the others. Bottom Panel: A comparison of the effective temperatures for the same stars shown in the top panel. APOGEE stars and the Arcturus result from Smith et al. (2013) are not shown given that we adopted their effective temperatures.}
  \label{fig:Fe-literature}
\end{figure}

A comparison of the metallicities obtained for stars in common with other studies in the literature is presented in Figure \ref{fig:Fe-literature}; the metallicities from the other works are from both low- and high- resolution optical and infrared spectra. The agreement between the metallicities is good, showing reasonable scatter and just a marginal systematic difference in the results is obtained: $<\delta$([Fe/H])``This study - Others''$>$ = 0.05 $\pm$0.11, rms = 0.12. 

Furthermore, it is important to verify whether there are large and obvious systematic differences between the different metallicity scales derived here: one from Fe I lines at $\sim\lambda$2.3$\mu$m (for most of the target stars), another from optical Fe I lines at $\sim\lambda$6000\AA, and a third taken directly from the APOGEE DR14 data release, which is based on overall model fits of APOGEE spectra. The ``standard'' Arcturus provides a good reference star to test the different metallicity scales, which we note is the reference red giant star for the APOGEE survey. 

Arcturus is one of the stars included in the comparison presented in Figure \ref{fig:Fe-literature}: the metallicity obtained for Arcturus $<$[Fe/H]$>$= -0.53$\pm$0.06 (Table \ref{tab:parameters}) was derived from spectral synthesis of Fe I lines at $\lambda$2.3$\mu$m using the FTS Arcturus atlas (Hinkle, Wallace \& Livingston 1995). The derived metallicity compares well with the one by J\"onsson et al. (2014b; obtained for T$_{eff}$ = 4226 K and log g = 1.67) of  $<$[Fe/H]$>$=-0.62; the latter was derived from an equivalent width analysis of optical Fe I lines measured from NARVAL archival spectra (Auri\`ere 2003). The difference for the metallicity of Arcturus derived in the two studies of 0.09 dex is small and  does not represent significant systematic offsets given all the uncertainties in the abundance determinations. A comparison of our result with the metallicity for Arcturus from Smith et al. (2013), obtained using the APOGEE line list at 1.5 micron and analyzing the same FTS spectra (Hinkle, Wallace \& Livingston 1995) also indicates good agreement: [Fe/H] (This study - APOGEE) = -0.06 dex; we note that both determinations adopt the same stellar parameters for Arcturus (T$_{eff}$ = 4275 K; log g = 1.70; $/xi$= 1.85 km/s). Arcturus is the standard red giant in the APOGEE survey and this star, along with the Sun, is used to derive astrophysical $gf$-values for the transitions in the APOGEE region. The metallicity adopted in the computation of the astrophysical $gf$-values is also in agreement with the derived metallicity for Arcturus.

To further assess possible systematic differences between the optical and infrared metallicity scales we also analyzed an optical spectrum of the ``standard'' star Arcturus. We performed two calculations: 1) adopting the same model atmospheres used in the analysis of the infrared lines with parameters taken from Smith et al. (2013) and deriving Fe abundances and, 2) iterating to a fully consistent solution for the effective temperature, surface gravity (from PARAM 1.3) and microturbulent velocity based on the equivalent width measurements of Fe I lines, similar to the methodology used for the other stars with optical spectra. The metallicity for Arcturus obtained in case (1) was [Fe/H]= -0.59 $\pm$ 0.11 and the results obtained from the fully iterated run were: [Fe/H] = -0.50 +/- 0.09; $T\rm_{eff}$ = 4280 K; log $g$ = 1.64; $\xi$ = 1.60 km/s (case 2). When compared to the metallicity obtained from the infrared lines of [Fe/H] = -0.53 $\pm$ 0.06), the metallicity obtained for the iterated solution is very similar (a difference of 0.03 dex).
For all purposes, although not homogeneous, we can consider that the metallicity scales in this study do not show systematic differences that are significant.

A comparison of the effective temperatures for stars in common with other studies in the literature is shown in the bottom panel of Figure 8.  The agreement is good, with a mean difference and standard deviation of $\delta$T$_{\rm eff}$(``This study - Others'') = -25$\pm$74K (not including the most discrepant result).  There is not a significant offset or trend in comparison to other literature temperature scales.

\section{Discussion}

\subsection{Chemical Evolution and the Sources of Fluorine}

While Section 2 pointed out that HF is best studied in the cooler red giants (T$_{eff}$ $\le$ 4000K), due to the rapidly decreasing HF line strength with increasing temperatures, the task of finding suitably cool red giants at low metallicity is a challenge, due to the shift in the RGB towards higher T$_{eff}$, at a given luminosity, with decreasing metallicity.  Nonetheless, the abundance results presented in Table \ref{tab:abundance2} show that HF has been detected, with F abundances derived, in red giants spanning a range of Fe abundances from roughly solar down to values of [Fe/H]$\sim$-1.2.  Over this interval in the Fe abundance, the F abundances vary by about of a factor of 25; these are the first F abundance measurements that span such a large range of metallicity and provide the most detailed view to date of on how F abundances change with changing Fe abundances.

\subsubsection{The Observed Behavior of Fluorine}

\begin{figure}[t!]
  \centering
  \includegraphics[width=\textwidth]{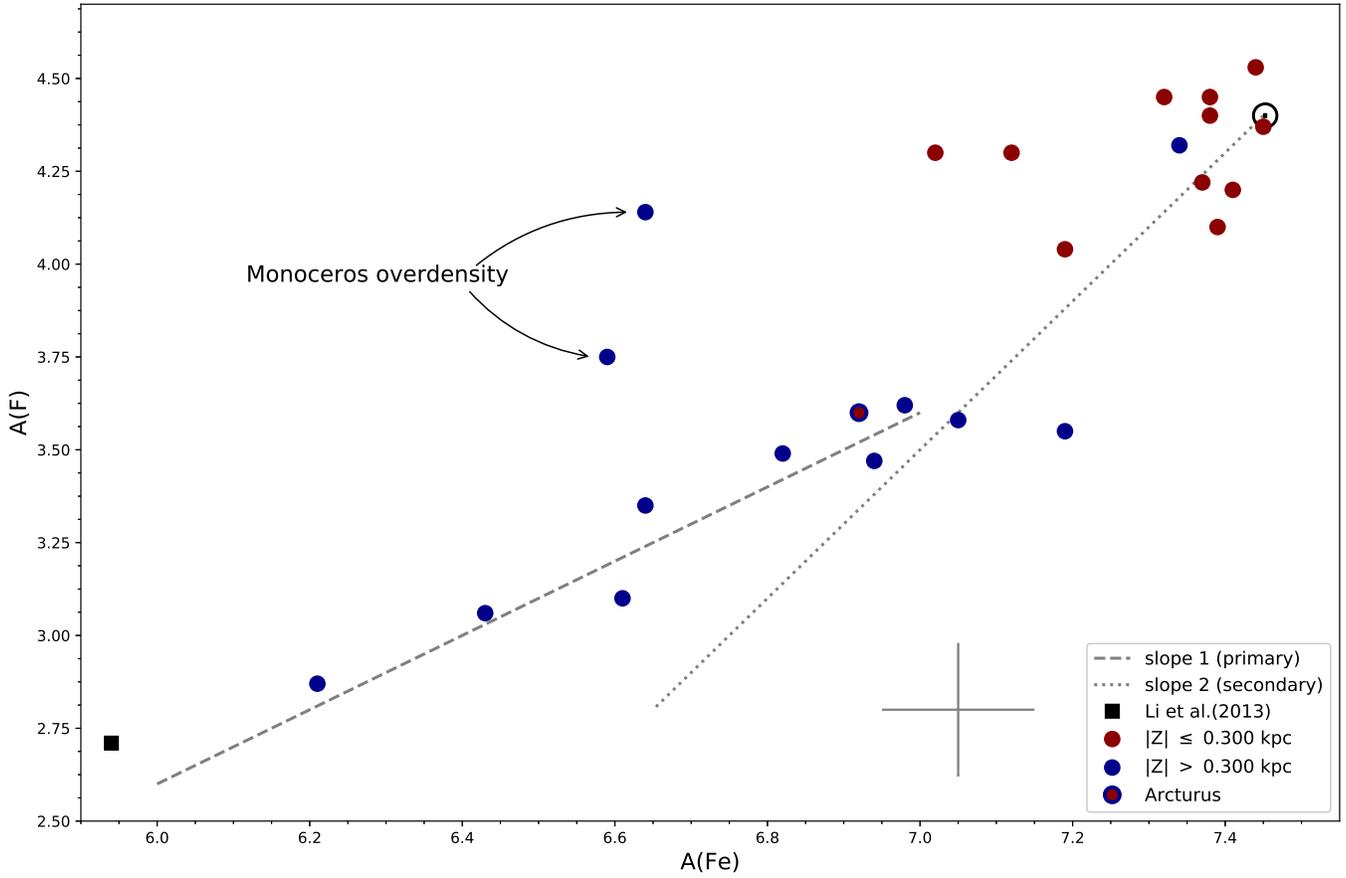}
  \caption{The behavior of the fluorine abundance with metallicity obtained for the studied stars. Dashed lines representing pure primary and secondary behaviors for the change of fluorine with metallicity are also shown. The stars are segregated according to their distances, Z, from the Galactic mid-plane. The blue circles correspond to probable thick disk/halo stars and the red circles to probable thin disk stars. Two stars identified as probable members of the Monoceros over-density are marked. A representative error bar is shown.}
  \label{fig:Fe_vs_F-2}
\end{figure}

Figure \ref{fig:Fe_vs_F-2} plots F versus Fe abundances for the red giants studied here; included is one low-metallicity red giant from Li et al.(2013) where HF was detected and analyzed.  (Li et al. detected HF in a second low-metallicity red giant, it is a CH star and thus has likely had its primoridal F abundance increased by He-burning and neutron captures (e.g., Jorissen et al. 1992) and in a third star exhibiting abundance patterns of dwarf spheroidals).  The cloud of points surrounding the solar abundances are the bright, nearby late-K and M giants that were observed with the KPNO 4m-FTS and these abundances cluster around the solar value or slightly below, with the size of the cloud roughly corresponding to the uncertainties in our derived abundances.
As A(Fe) decreases below about 7.2, the fluorine abundance is found to rapidly decrease by about $\sim$0.5 dex over a very narrow range in Fe (less than $\sim$0.1 dex).  At lower Fe abundances, the majority of red giants have F abundances that decrease linearly in the logarithmic A(F)-A(Fe) plane, with a slope of very nearly 1.0.  This type of chemical evolution demonstrates that F behaves as a primary product of nucleosynthesis relative to Fe; the long-dashed line in Figure \ref{fig:Fe_vs_F-2} has a slope of 1.0 and is set to pass approximately through the low-metallicity points to demonstrate primary chemical evolution of F with Fe (we note that this is not a fit to the points).  The dotted line that originates at the solar abundance with decreasing F and Fe abundances has a slope of 2.0 and is plotted to demonstrate how secondary chemical evolution with Fe would behave.  Although at the higher metallicities the chemical evolution of F may be complex, its behavior at Fe abundances below about -0.5 dex solar displays a clear primary behavior with Fe.  

One further item to note in Figure \ref{fig:Fe_vs_F-2} is the position of two metal-poor red giants that fall well above the A(F)-A(Fe) line that defines most of the metal-poor stars; these two red giants have been identified, based upon their Galactic positions, space motion, and peculiar chemistry, to be probable members of the Monoceros overdensity, which is a feature in the outer Galaxy and is probably an over-density structure in the outer disk, or, as initially proposed by Newberg et al. (2002), related to an accreted dwarf galaxy. The stellar members of this structure display peculiar chemical abundance distributions (Chou et al. 2010) and, thus, may not display the same F-Fe abundance patterns of more nearby Galactic populations. 
Alternatively, this abundance pattern could be representative of the outer thin disk, if Monoceros is a structure of the flared thin disk. 

Figure \ref{fig:Fe_vs_F2Fe} is similar to Figure \ref{fig:Fe_vs_F-2}, but uses two panels to compare the derived values of [F/Fe] versus [Fe/H] (including results from J\"onsson et al. (2014b; 2017), Pilachowski \& Pace (2015) and Li et al. (2013)) to predictions from a variety of models.
The top panel shows only the derived stellar values of [F/Fe], while the bottom panel overlays the various chemical evolution models over the observed abundances.  
In the near-solar metallicity regime, all of the various abundance results roughly overlap, although exhibit scatter that can be at least partially explained by the internal errors and possible systematic in the different studies, 
or, it is also possible that part of the scatter may be real.
At values of [Fe/H]$<~$-0.4, [F/Fe] displays a roughly constant value, within the uncertainties, which is the signature of F behaving as a primary product relative to Fe (as noted in Figure \ref{fig:Fe_vs_F-2}). The flat dashed-line in the top panel of Figure \ref{fig:Fe_vs_F2Fe} is the average value of [F/Fe] for the red giants from this study having Fe abundances of less than [Fe/H]$\sim$ -0.3 and values of [F/Fe]$<$-0.2: this mean value is [F/Fe]=-0.36$\pm$0.10.
The dashed line connecting the solar value to the plateau line has a slope of 1.0, illustrating secondary F production with respect to Fe.  Note that the results from J\"onsson et al. (2017) for the slightly metal-poor stars in their sample scatter about this secondary line.  
The observationally-derived behavior of [F/Fe] as a function of [Fe/H] suggests a primary production of F at low metallicities with a near-constant subsolar value of [F/Fe]$\sim$-0.36, with a secondary source possibly driving a significant increase in the F abundance over a small interval in the Fe abundance. 

\subsubsection{Comparisons with Chemical Evolution Models}

\begin{figure}[t!]
  \centering
  \includegraphics[width=\textwidth]{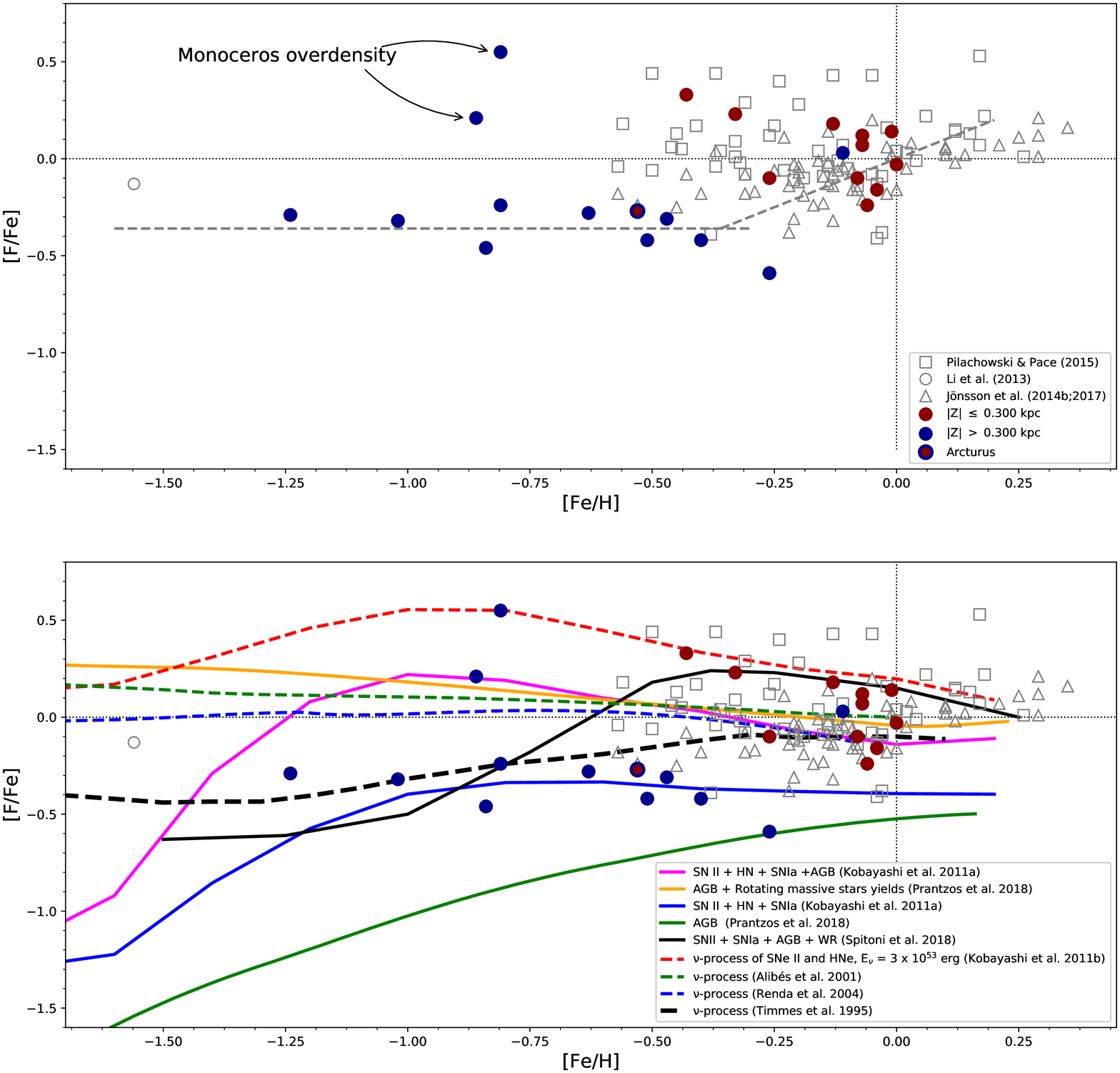}
  \caption{The chemical evolution of fluorine viewed as [F/Fe] versus [Fe/H]. The blue circles represent stars with distances from the mid-plane $|Z|>$ 300 pc, corresponding to the geometric thick disk / halo, while the red circles correspond to probable thin disk stars ($|Z|<$ 300 pc). Other fluorine abundance results from the literature are also shown as open symbols. Several chemical evolution models from the literature are also shown (bottom panel; see the figure key).}
  \label{fig:Fe_vs_F2Fe}
\end{figure}

The abundances of F and Fe illustrated in Figure \ref{fig:Fe_vs_F2Fe} can be used to evaluate contributions from possible sources for the nucleosynthesis of $^{19}$F based on chemical evolution model predictions by Timmes et al. (1995), Alib\'es et al. (2001), Renda et al. (2004), Kobayashi et al. (2011a; 2011b), Spitoni et al. (2018), and Prantzos et al. (2018); shown in the bottom panel of Figure \ref{fig:Fe_vs_F2Fe}.  Likely sources to consider are both neutrino nucleosynthesis (or the ``$\nu$-process''; Woosley et al. 1990; Timmes et al. 1995; Alib\'es et al. 2001; Kobayashi et al. 2011a; 2011b) and AGB stars (Jorissen et al. 1992; Forestini et al. 1992; Abia et al. 2009; Cristallo et al. 2014; Spitoni et al.2018).  
Massive stars have also been proposed as a significant source of fluorine (e.g., Kobayahsi et al. 2011a), including mass loss in Wolf-Rayet stars (Meynet \& Arnould 2000; Spitoni et al. 2018), as well as rapidly rotating metal-poor massive stars (Prantzos et al. 2018).  
Several chemical evolution models are shown in the bottom panel of Figure \ref{fig:Fe_vs_F2Fe} as comparisons to the observations/derived abundances.

Focusing first on the predictions of [F/Fe] versus [Fe/H] from the four models including neutrino nucleosynthesis from Timmes et al. (1995), Alib\'es et al. (2001), Renda et al. (2004) and Kobayashi et al. (2011b), it is noted that they display significant differences and highlight uncertainties in such input data as reaction rates, the neutrino spectrum, or details of the SN II explosion (e.g., Heger et al. 2005; Sieverding et al. 2018; 2019), or modelling the chemical evolution of the Milky Way.  The $\nu$-process models result in rather flat distributions of [F/Fe] versus [Fe/H], which describe the observed near-primary trend in the observed points at low metallicity.
However, it is clear that no single model fits the observed abundances, although the early model from Timmes et al. (1995) fits the low-metallicity ``plateau'' fairly well, as well as the raise to near-solar [F/Fe] values at solar metallicity.

The bottom panel of Figure \ref{fig:Fe_vs_F2Fe} also illustrates models with other sources of $^{19}$F that predict a near-primary behavior with Fe that results from pre-SN mass loss in low-metallicity rapidly-rotating massive stars during He-burning, as modelled recently by Prantzos et al. (2018); this is shown by the orange curve.  The rotating massive stars modelled by Prantzos et al. (2018) synthesize $^{19}$F by initial $\alpha$-captures onto $^{14}$N: $^{14}$N($\alpha$,$\gamma$)$^{18}$F($\beta^{+}$,$\nu$)$^{18}$O(p,$\alpha$)$^{15}$N($\alpha$,$\gamma$)$^{19}$F (e.g., Goriely et al. 1990; Jorissen et al. 1992).  This series of reactions requires a proton capture in a nominally He-burning region, with the protons released via $^{14}$N(n,p)$^{14}$C and the initiating neutrons from $^{13}$C($\alpha$,n)$^{16}$O, thus the similarity to F-production in the He-burning thermal pulses of AGB stars.  This set of reactions in the low-metallicity rapidly-rotating massive stars leads to a near-primary behavior of F with respect to Fe as the $^{14}$N arises from $^{12}$C produced by He-burning in the massive star itself, whereas in AGB stars, the $^{}$N is synthesized only by proton captures onto the initial abundance of $^{12}$C (the classic CN-cycle), thus the secondary-like behavior of F from an AGB source.  The model from Prantzos et al. (2018) produces values of [F/Fe] that are nearly constant over the metallicity range spanned in the figure (similar to the behavior of the $\nu$-process models).  This model results in a near-solar (or thin disk) value of [F/Fe] at [Fe/H]$\sim$0.0, but over-produces fluorine below [Fe/H] $\sim$ -0.5 when compared to the thick disk/halo stars.  Prantzos et al. (2018) point out that the efficiency of rotation-driven mixing cannot be determined without external calibration.  The observed behavior of [F/Fe] at low metallicity may help to improve model predictions from this process, although a downward adjustment to the yields at low metallicity might then lead to under-production of fluorine at solar metallicity.  In addition, model F7 from Spitoni et al. (2018) also produces a near-primary behavior of F with Fe over the range of [Fe/H]$<$-1.5 up to [Fe/H]$\sim$-1.0.  The primary behavior of F here is driven by mass loss in rotating Wolf-Rayet stars.

Figure \ref{fig:Fe_vs_F2Fe} (bottom panel) also includes curves of [F/Fe] versus [Fe/H] for two chemical evolution models from Kobayashi et al. (2011a) that include fluorine contributions from massive star supernovae (blue curve), as well as a model that adds AGB stars to the massive SN yields (the magenta curve).
We note that the model with $^{19}$F synthesis from the massive stars results from their He-burning phase of stellar evolution and that this model also fits the plateau in [F/Fe] at low metallicity (blue curve).  We note also that, although the rapid increase in [F/Fe] from AGB stars in the Kobayashi et al. (2011a) model at [Fe/H]$\sim$-1.50 occurs at a lower metallicity than is observed, this particular model indicates that AGB stars can drive the rise in $^{19}$F abundances to solar-like values of [F/Fe] due to the various reaction chains that lead to the production of $^{19}$F beginning with alpha captures onto $^{14}$N.  As a comparison of models, the green curve in Figure \ref{fig:Fe_vs_F2Fe} illustrates the behavior of the AGB source from Prantzos et al. (2018), which results in lower AGB yields when compared to Kobayashi et al. (2011a), due to, primarily, differences in reaction rates. 

There remain differences and uncertainties in the models, but a comparison of derived fluorine abundances with models of chemical evolution suggests that the tightest constraints on potential sources of F will be provided by observations of a larger number of metal-poor stars, as well as probing fluorine abundances across a larger range of galactic environments and populations.

\subsection{The Fluorine Abundance Gradient}

Metallicity gradients are a common feature in galaxies and our own Galaxy is no exception; it is well known that metallicities in galaxy disks decrease with increasing distance from the galactic center. In the Milky Way, radial metallicity gradients are observed in several stellar populations, such as H II regions (e.g., Esteban \& Garc\'ia-Rojas 2018), B stars (e.g., Bragan\c{c}a et al. 2019; Daflon \& Cunha 2004), Cepheids (e.g., Genovali et al. 2014; Lemasle et al. 2008), open clusters (e.g., Yong et al. 2012; Magrini et al. 2017; Donor et al. 2018; Carrera et al. 2019), and planetary nebulae (e.g., Stanghellini \& Haywood 2018). In addition, results from recent spectroscopic surveys confirm the existence of chemical abundance gradients for field low-mass stars, indicating that gradients also vary with distance from the galactic plane (e.g., Cheng et al. 2012; Hayden et al. 2015) and age of the population (e.g., Bergemann et al. 2014; Anders et al. 2017), while radial mixing, via heating and/or migration, affects the radial abundance gradients (Sch\"onrich \& Binney 2009; Minchev et al. 2013; Kubryk et al. 2015).
There is a vast literature on radial gradients for several elements in the Milky Way but the fluorine abundance gradient has not been yet probed.

In this study we have the opportunity to investigate the variation of the fluorine abundance (along with iron) as a function of galactocentric distance. Our sample contains stars between roughly R$_g$ $\sim$ 8.7 -- 13 kpc, with one star in the inner Galactic region at R$_g$ $\sim$6 kpc. A significant fraction of our targets are in the solar neighborhood, having galactocentric distances close to the solar value (R$_{g_\odot}$ = 8.33 $\pm$ 0.35 kpc, Gillessen et al. 2009).
Given the goal of probing fluorine at low metallicities, roughly half of the targets were selected to be both cool and metal-poor, possibly being from the thick disk or halo, from known over-densities, or stellar sub-structures. 
In fact, two of the targeted stars (2MASS01051470+4958078 and 2MASS22341156+4425220) lie at $|Z|$ $\sim$ 1.2 kpc from the Galactic plane and appear be members of the halo, with rotational velocities $V_{\phi} \sim 0$ km s$^{-1}$ (calculated using Gaia DR2 combined with APOGEE DR14 radial velocities), but are relatively metal-rich for the halo with [Fe/H] $\sim -1$, suggesting that they may be members of the Gaia-Enceladus or Gaia-Sausage accretion event \citep{belokurov2018,helmi2018}.  Examining the APOGEE DR14 \citep{holtzman2018} chemical abundances for these stars, they have low [X/Fe] for (C$+$N), Mg, Al, and Ni, consistent with the chemical abundance profile of this accreted population in the Milky Way halo \citep{hayes2018}. 

Two other stars in our sample (2MASS02431985+5227501 and 2MASS04224371+1729196) are notable due to their high [F/Fe] for their relatively low metallicities of [Fe/H] $\sim$ -0.8 (see Figures \ref{fig:Fe_vs_F-2} and \ref{fig:Fe_vs_F2Fe}). These two stars lie below the Milky Way plane by 1-2 kpc at Galactocentric radii $R_{GC} \gtrsim 11$ kpc, but are rotating in the same direction as the disk, with rotational velocities around $V_{\phi} \sim 230$ km s$^{-1}$.  This suggests that these stars are perhaps members of the Monoceros overdensity, which has been reported as a metal-poor overdensity moving with velocities similar to the disk, but lying below the nominal disk mid-plane by 1-3 kpc at these Galactocentric radii \citep{rochapinto2003,crane2003,ibata2003,morganson2016}.  Additionally, these stars have nearly solar $\alpha$-element abundances consistent with the chemistry of the Monoceros Ring \citep{chou2010}.

\begin{figure}[t!]
  \centering
  \includegraphics[width=\textwidth]{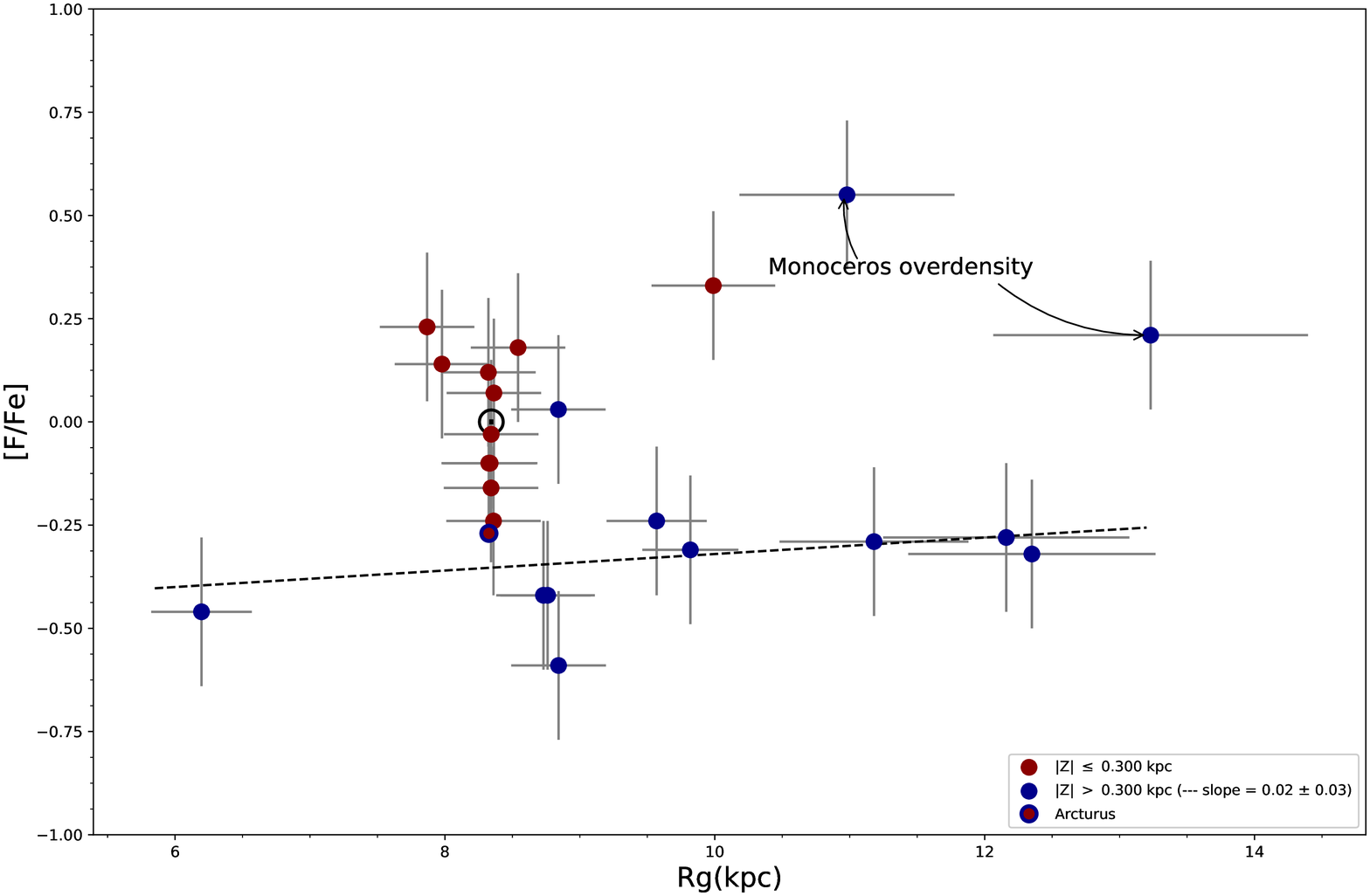}
  \caption{The [F/Fe] abundances as a function of the galactocentric distance, R$_g$. The blue circles represent probable thick disk/halo stars and the red circles thin disk stars. The dashed line represents the best fit slope to the thick disk/halo stars, as they span a significant range in R$_g$, excluding the two stars that are identified as probable members of the Monoceros over-density. The distances and uncertainties are based on Gaia DR2 and were taken from Bailer-Jones et al. (2018).}
  \label{fig:F2Fe_vs_Rg}
\end{figure}

The [F/Fe] ratio versus R$_g$ for the studied sample, is shown in Figure \ref{fig:F2Fe_vs_Rg}. Most of the targets have Gaia DR2 distances with estimated errors of the order of 0.3 -- 0.4 kpc (Bailer-Jones et al. 2018; Table \ref{tab:observations}). 
Here again we segregate the targets based on a purely geometric definition of the thin/thick disk: stars having $|Z|>$300 pc taken as probable thick disk/halo population (filled blue circles) and $|Z|<$ 300 pc as probable thin disk stars (filled red circles).
Although having $|Z|<$ 300 pc, Arcturus was not included in the thin disk sample given that it is known to be a thick disk star (Ramirez et al. 2007).
It is clear that the sample of thin disk stars in our study does not cover a range in galactocentric distances large enough to define a gradient. For the solar neighborhood stars the [F/Fe] abundances cluster around the solar [F/Fe] value with a mean $<$[F/Fe]$>$= 0.01 $\pm$ 0.15.
Only one probable thin disk star (2MASS07313775-2818395) is located $\sim$2 kpc beyond the solar vicinity and has an elevated [F/Fe] ratio when compared to the solar average for the solar neighborhood. 

Focusing on the sample of probable thick disk / halo stars, the gradient of [F/Fe] for this population has a shallow slope of 0.02$\pm$0.03 dex / kpc, and given its error, this is consistent with a flat behavior for [F/Fe] with galactocentric distance between $\sim$ 6 - 13 kpc. 
The [F/Fe] ratio for these stars, all having $\vert$Z$\vert$ $>$ 300 pc above the Galactic plane, is subsolar, indicating a quasi-plateau corresponding to $\sim$-0.3 dex. As discussed in the previous section, such a floor in [F/Fe] likely corresponds to the production of fluorine in SN II via neutrino nucleosynthesis (Figure \ref{fig:Fe_vs_F2Fe}), while the thin disk stars between $\sim$ 8 -- 10 kpc would potentially follow a sequence with a steeper positive slope with R$_g$, having [F/Fe] from roughly solar to $\sim$+0.3 dex, although this is based on the abundances obtained for one single more distant thin disk star in our sample and 
a pure speculation at this point.

A relatively flat abundance gradient for the thick disk is in general agreement with other results from the literature (e.g., Boeche et al. 2013, Cheng et al. 2012; Mikolaitis et al. 2014; Hayden et al. 2015; Weinberg et al. 2019). 
We note, however, the very discrepant positions in the plane [F/Fe] - R$_g$ of those targets tagged as being possible members of the Monoceros over-density. These two stars have significantly higher fluorine abundances for their metallicities, among the highest [F/Fe] in our sample, with a mean [F/Fe] = +0.38 dex (see also their location in Figures \ref{fig:Fe_vs_F-2} and \ref{fig:Fe_vs_F2Fe}). Compared with the other stars in the outer disk sample, their F/Fe abundances are roughly 1 dex higher than those for stars at similar $R_g$. Such discrepancy is well beyond the expected uncertainties in the abundance determinations that have been done homogeneously.
Given the clear inconsistent abundances obtained for these two Monoceros targets, these stars were not included in the derivation of the radial gradients discussed above. 
The very distinct fluorine abundance behavior for the two possible Monoceros targets when compared with the radial gradient obtained for other target stars away from the mid-plane ($|Z|>$ 300 pc) adds to the on-going debate about the origin of the Monoceros over-density. Given their high fluorine content, and the fact that in the outer galaxy the chemical thin disk flares (Minchev et al. 2015, 2017; Bovy 2016; Mackereth et al. 2017) it would be tempting to speculate (just based on their abundances) that the abundance pattern in these stars could represent an extension of the fluorine gradient in the thin disk. 
However, given the small number of stars analyzed and the lack of outer thin disk stars in our sample, this speculation should be viewed with extreme caution. It is clear, however, that the fluorine abundances for these two Monoceros stars do not match those of the thick disk/halo stars.

\section{Conclusions}


Fluorine and iron abundances are presented for a sample of Galactic red giant stars and are used to map the behavior of [F/Fe] as a function of metallicity, spanning the largest range in metallicity to date ([Fe/H] $\sim$ -1.3 to +0.0) and providing insights into the nucleosynthesis of $^{19}$F.

Two regimes in the behavior of [F/Fe] are found: 
1) at low metallicities (-1.3$<$[Fe/H]$<$-0.4), fluorine abundances behave as a primary element with respect to iron; 2) at high metallicities ([Fe/H]$>$-0.4 to $\sim$0.0), the F abundances are near-secondary with Fe.  
No single chemical evolution model satisfactorily describes the observed behavior of the derived fluorine abundances as a function of metallicity.
A primary-like behavior of the fluorine abundance with the Fe abundance, over the range of [Fe/H]$\sim$ -1.5 to -0.4, is predicted by models including only fluorine production in neutrino nucleosynthesis (Timmes et al. 1995; Alib\'es et al. 2001; Renda et al. 2004), as well as rapidly-rotating low-metallicity massive stars (Prantzos et al. 2018), although the quantitative values of the [F/Fe] plateau from different models remains disparate and uncertain. 
The early neutrino nucleosynthesis model by Timmes et al. (1995) roughly describes the observed [F/Fe] plateau at low metallicities and the rise to [F/Fe] to near solar at roughly solar metallicities.
Models including AGBs as fluorine sources (Prantzos et al. 2018 and Kobayashi et al. 2011a) show large discrepancies due to uncertainties in the reaction rates and yields.

This is the first study to measure fluorine abundance gradients in the Milky Way. 
The [F/Fe] gradient for thick disk/halo stars (defined as those having $|$Z$|>$ 300 pc) is shallow with a slope of 0.02 $\pm$ 0.03 dex / kpc between  R$_g$ $\sim$ 6 - 13 kpc. For this population the [F/Fe] ratio is subsolar, while for thin disk stars in the solar neighborhood (Rg $\sim$ 8 kpc) the [F/Fe] ratio is roughly solar. Our sample contains only one thin disk star ($|$Z$|<$ 300 pc) beyond the solar neighborhood (R$_g$ $\sim$ 10 kpc) and it has a higher value of [F/Fe] $\sim$ +0.3 dex.

Further progress in understanding the chemical evolution of fluorine would be helped by additional measurements of fluorine abundances at low metallicities, as well as measuring its abundance in the outer disk regions of the Milky Way.

\acknowledgments
\section{Acknowledgments}

We thank the referee for suggestions that improved the paper. We thank Nikos Prantzos for discussions about the models.
R.G. and K.C. thank Gustavo Bragan\c{c}a and Simone Daflon for discussions about gradients. 
H.J. acknowledges support from the Crafoord Foundation, Stiftelsen Olle Engkvist Byggm{\"a}stare, and Ruth och Nils-Erik Stenb{\"a}cks stiftelse. 
CA ackowledges the Spanish grants AYA2015-63588-P and
PGC2018-095317-B-C21
within the European Founds for Regional Development (FEDER).

{\it Facilities: {Gemini Observatory}, {Kitt Peak National Observatory}, {NASA Infrared Telescope Facility}}.
\software{IRAF (Tody 1986, Tody 1993), Turbospectrum (Alvarez \& Plez 1998; Plez 2012), MOOG (Sneden et al. 2012).}


\clearpage

\begin{sidewaystable}
\centering
\scalefont{0.65}
\caption{Sample Stars \& Observations}
\begin{tabular}{lccccccccccc}
\hhline{============}
                       & Spectrograph        & Spectrograph  & $\pi$             & d                   &          &          &           &          &           & R$_g$           & Z \\
Star                   & (infrared)          & (Optical)     & (mas)             & (pc)                & E(B-V)   & A$_V$    & V         & J        & K         & (kpc)           & (kpc)                  \\ \hline
Arcturus               & FTS/KPNO 4-m        & ...           & 88.8$\pm$0.54 (c) & 11.26$\pm$0.070 (d) & ...      & ...      & -0.05 (e) & ...      & -3.00 (e) & 8.326$\pm$0.350 & 0.01                   \\
10 Dra                 & FTS/KPNO 4-m        & ...           & 8.21$\pm$0.26 (a) & 121.6$\pm$4.100 (b) & ...      & ...      &  4.66 (e) & ...      & -0.17 (e) & 8.360$\pm$0.350 & 0.09                   \\
41 Com                 & FTS/KPNO 4-m        & ...           & 8.60$\pm$0.37 (a) & 116.3$\pm$5.400 (b) & ...      & ...      &  4.82 (e) & ...      &  1.20 (e) & 8.325$\pm$0.350 & 0.12                   \\
$\beta$ And            & FTS/KPNO 4-m        & ...           & 16.5$\pm$0.56 (c) & 60.53$\pm$1.990 (d) & ...      & ...      &  2.05 (e) & ...      & -1.87 (e) & 8.363$\pm$0.350 & -0.03                  \\
$\beta$ Peg            & FTS/KPNO 4-m        & ...           & 16.6$\pm$0.15 (c) & 60.10$\pm$0.540 (d) & ...      & ...      &  2.42 (e) & ...      & -2.22 (e) & 8.335$\pm$0.350 & -0.03                  \\
$\beta$ UMi            & FTS/KPNO 4-m        & ...           & 24.9$\pm$0.12 (c) & 40.14$\pm$0.190 (d) & ...      & ...      &  2.08 (e) & ...      & -1.39 (e) & 8.342$\pm$0.350 & 0.03                   \\
$\gamma$ Dra           & FTS/KPNO 4-m        & ...           & 21.1$\pm$0.10 (c) & 47.30$\pm$0.220 (d) & ...      & ...      &  2.22 (e) & ...      & -1.35 (e) & 8.322$\pm$0.350 & 0.02                   \\
$\mu$ Leo              & FTS/KPNO 4-m        & ...           & 30.7$\pm$0.42 (a) & 32.60$\pm$0.470 (b) & ...      & ...      &  3.88 (e) & ...      &  1.22 (e) & 8.349$\pm$0.350 & 0.03                   \\
$\omega$ Vir           & FTS/KPNO 4-m        & ...           & 7.32$\pm$0.42 (a) & 136.8$\pm$8.500 (b) & ...      & ...      &  5.36 (e) & ...      & -0.30 (e) & 8.343$\pm$0.350 & 0.12                   \\
BD$+$06$^o$2063        & FTS/KPNO 4-m        & ...           & 1.25$\pm$0.06 (a) & 780.2$\pm$36.60 (b) & ...      & ...      & ...       & ...      & ...       & 8.841$\pm$0.350 & 0.39                   \\
BD$+$16$^o$3426        & FTS/KPNO 4-m        & ...           & 1.91$\pm$0.05 (a) & 515.0$\pm$14.60 (b) & ...      & 0.39 (f) &  8.03 (g) & ...      &  2.19 (h) & 7.978$\pm$0.350 & 0.14                   \\
HD 96360               & FTS/KPNO 4-m        & ...           & 2.37$\pm$0.04 (a) & 416.6$\pm$7.000 (b) & ...      & ...      & ...       & ...      & ...       & 8.542$\pm$0.350 & 0.30                   \\
HD 189581              & FTS/KPNO 4-m        & ...           & 1.49$\pm$0.07 (a) & 662.1$\pm$0.300 (b) & ...      & 0.46 (f) &  8.45 (g) & ...      &  2.71 (h) & 7.868$\pm$0.350 & -0.17                  \\
TYC 5880-423-1         & Phoenix/Gemini-S    & ...           & 0.38$\pm$0.04 (a) & 2375.$\pm$234.9 (b) & 0.03 (i) & 0.09 (j) & 10.97 (k) & ...      &  7.46 (m) & 9.569$\pm$0.371 & -1.86                  \\
UCAC2 22625431         & Phoenix/Gemini-S    & ...           & 0.13$\pm$0.04 (a) & 5232.$\pm$1038. (b) & 0.06 (i) & 0.18 (j) & 12.28 (l) & ...      &  8.10 (m) & 12.16$\pm$0.912 & -1.26                  \\
2M06232713$-$3342412 & Phoenix/Gemini-S      & ...           & 0.33$\pm$0.02 (a) & 2730.$\pm$166.1 (b) & 0.04 (i) & 0.12 (j) & ...       & 8.84 (m) &  7.76 (m) & 9.819$\pm$0.356 & -0.94                  \\
2M07313775$-$2818395 & Phoenix/Gemini-S      & ...           & 0.32$\pm$0.05 (a) & 2885.$\pm$459.8 (b) & 0.20 (i) & 0.61 (j) & ...       & 8.37 (m) &  7.09 (m) & 9.989$\pm$0.458 & -0.23                  \\
HD 19697               & Phoenix/KPNO 2.1-m  & SES/MCD 2.1-m & 1.50$\pm$0.06 (a) & 653.5$\pm$24.70 (b) & ...      & ...      & ...       & ...      & ...       & 8.842$\pm$0.351 & -0.39                  \\
HD 20305               & Phoenix/KPNO 2.1-m  & SES/MCD 2.1-m & 1.81$\pm$0.08 (a) & 544.2$\pm$23.70 (b) & ...      & ...      & ...       & ...      & ...       & 8.761$\pm$0.350 & -0.32                  \\
HD 28085               & Phoenix/KPNO 2.1-m  & SES/MCD 2.1-m & 2.57$\pm$0.04 (a) & 384.9$\pm$6.160 (b) & ...      & ...      & ...       & ...      & ...       & 8.688$\pm$0.350 & -0.14                  \\
HD 90862               & Phoenix/KPNO 2.1-m  & SES/MCD 2.1-m & 1.02$\pm$0.05 (a) & 950.2$\pm$49.70 (b) & ...      & ...      & ...       & ...      & ...       & 8.730$\pm$0.350 & -0.79                  \\
2M01051470$+$4958078 & iShell/IRTF + APOGEE  &  ...          & 0.10$\pm$0.03 (a) & 5640.$\pm$1052. (b) & ...      & ...      & ...       & ...      & ...       & 12.35$\pm$0.916 & -1.25                  \\
2M02431985$+$5227501 & iShell/IRTF + APOGEE  &  ...          & 0.11$\pm$0.04 (a) & 5768.$\pm$1231. (b) & ...      & ...      & ...       & ...      & ...       & 13.23$\pm$1.166 & -0.68                  \\
2M04224371$+$1729196 & iShell/IRTF + APOGEE  &  ...          & 0.20$\pm$0.10 (a) & 2865.$\pm$773.3 (b) & ...      & ...      & ...       & ...      & ...       & 10.98$\pm$0.797 & -1.08                  \\
2M20410375$+$0001223 & iShell/IRTF + APOGEE  &  ...          & 0.16$\pm$0.04 (a) & 4696.$\pm$1003. (b) & ...      & ...      & ...       & ...      & ...       & 6.196$\pm$0.373 & -1.92                  \\
2M22341156$+$4425220 & iShell/IRTF + APOGEE  & ...           & 0.10$\pm$0.02 (a) & 6448.$\pm$963.1 (b) & ...      & ...      & ...       & ...      & ...       & 11.18$\pm$0.700 & -1.33                  \\ \hline
\end{tabular}
\begin{tablenotes}
\item \textbf{Notes}: (a) Gaia DR2; (b) Bailer-Jones et al. (2018); (c) Hipparcos parallaxes; (d) Distances determined from inverted Hipparcos parallaxes; (e) Ducati (2002); (f) Chen et al. (1998a); (g) Hog (2000); (h) Chen et al. (1998b); (i) Green et al. (2018); (j) A$_V$ using R$_V$ = 3.07 from McCall (2004); (k) Munari et al. (2014); (l) Zacharias et al. (2012); (m) K obtained with Carpenter (2001) transformation from (K$_S$)$_{2MASS}$.
\end{tablenotes}
\label{tab:observations}
\end{sidewaystable}

\clearpage

\begin{table}
\centering
\caption{Stellar Parameters and abundances}
\begin{tabular}{lcccccc}
\hhline{=======}
                       & T$_{eff}$ & $\log$g & $\xi$               &          &           &        \\
Star                   & (K)       & [cgs]   & (km$\cdot$s$^{-1}$) &  A(Fe) & A(F)        & [F/Fe] \\ \hline
Arcturus*              & 4275      & 1.70    & 1.85 &  6.92 $\pm$ 0.06  & 3.63 $\pm$ 0.12 & -0.27  \\
10 Dra                 & 3559      & 0.69    & 1.90 &  7.39 $\pm$ 0.05  & 4.10 $\pm$ 0.07 & -0.24 \\
41 Com                 & 3917      & 1.30    & 1.75 &  7.37 $\pm$ 0.06  & 4.22 $\pm$ 0.07 & -0.10 \\
$\beta$ And            & 3801      & 1.00    & 1.80 &  7.38 $\pm$ 0.04  & 4.40 $\pm$ 0.00 &  0.07 \\
$\beta$ Peg            & 3599      & 0.75    & 1.90 &  7.19 $\pm$ 0.08  & 4.04 $\pm$ 0.05 & -0.10 \\
$\beta$ UMi            & 3984      & 1.32    & 1.70 &  7.41 $\pm$ 0.05  & 4.20 $\pm$ 0.05 & -0.16 \\
$\gamma$ Dra           & 3939      & 1.21    & 1.75 &  7.38 $\pm$ 0.05  & 4.45            &  0.12 \\
$\mu$ Leo              & 4486      & 2.48    & 1.45 &  7.75 $\pm$ 0.13  & ...             &  ...  \\
$\omega$ Vir           & 3418      & 0.49    & 1.95 &  7.45 $\pm$ 0.07  & 4.37 $\pm$ 0.10 & -0.03 \\
BD$+$06$^o$2063**      & 3550      & 0.70    & 1.90 & 7.34 $\pm$ 0.09  & 4.32 $\pm$ 0.01  &  0.03 \\
BD$+$16$^o$3426        & 3489      & 0.64    & 1.90 & 7.44 $\pm$ 0.09  & 4.53 $\pm$ 0.08  &  0.14 \\
HD 96360**             & 3550      & 0.72    & 1.90 & 7.32 $\pm$ 0.08  & 4.45 $\pm$ 0.02  &  0.18 \\
HD 189581              & 3406      & 0.51    & 1.95 & 7.12 $\pm$ 0.10  & 4.30 $\pm$ 0.05  &  0.23 \\
HD 19697               & 3850      & 1.23    & 1.39 & 7.19 $\pm$ 0.21  & 3.55             & -0.59 \\
HD 20305               & 4075      & 1.45    & 1.67 & 7.05 $\pm$ 0.10  & 3.58             & -0.42 \\
HD 28085               & 4850      & 2.33    & 1.65 & 7.41 $\pm$ 0.11  & ...              &  ...  \\
HD 90862               & 3916      & 0.95    & 1.75 & 6.94 $\pm$ 0.16  & 3.47             & -0.42 \\
TYC 5880-423-1         & 3840      & 0.84    & 1.77 & 6.64             & 3.35             & -0.24 \\
UCAC2 22625431         & 3631      & 0.32    & 1.86 & 6.82             & 3.49             & -0.28 \\
2M06232713$-$3342412 & 3860        & 0.95    & 1.74 & 6.98             & 3.62             & -0.31 \\
2M07313775$-$2818395 & 3800        & 0.82    & 1.78 & 7.02             & 4.30             &  0.33 \\
2M01051470$+$4958078 & 3872        & 0.50    & 1.86 & 6.43$\pm$0.03    & 3.06             & -0.32 \\
2M02431985$+$5227501 & 3673        & 0.28    & 1.88 & 6.59             & 3.75             &  0.21 \\
2M04224371$+$1729196 & 3627        & 0.21    & 1.88 & 6.64$\pm$0.03    & 4.14             &  0.55 \\
2M20410375$+$0001223 & 3738        & 0.39    & 1.85 & 6.61$\pm$0.05    & 3.10             & -0.46 \\
2M22341156$+$4425220 & 4016        & 0.52    & 1.79 & 6.21$\pm$0.02    & 2.87             & -0.29 \\ \hline
\hline
\end{tabular}
\begin{tablenotes}
\item \textbf{Notes}: Solar fluorine and iron abundances taken from Maiorca et al. (2014) and Grevesse et al. (2007), respectively. * Stellar parameters from Smith et al. (2013); ** Stellar parameters from Jorissen et al. (1992).
\end{tablenotes}
\label{tab:parameters}
\end{table}

\clearpage

\begin{table}
\centering
\caption{Infrared Fluorine and Iron lines}
\begin{tabular}{ccclrcc}
\hhline{=======}
Species /  & $\lambda _{air}$ &          & $\chi$        &             & $\Gamma_6$ & D$_\circ$ \\ 
Molecule   & (\AA)           & Line ID   & (eV)          & $\log$ $gf$ &          & (eV)      \\ \hline
H$^{19}$F  & 23358.329       & (1--0)R9  & 0.227 (a,b)   & -3.962(a)   & ...      & 5.869 (c) \\
           & 22957.938       & (1--0)R13 & 0.455 (a)     & -3.941(a)   & ...      & 5.869 (c) \\
           & 22886.733       & (1--0)R14 & 0.524 (a)     & -3.947(a)   & ...      & 5.869 (c) \\
           & 22826.862       & (1--0)R15 & 0.597 (a)     & -3.956(a)   & ...      & 5.869 (c) \\
           & 22778.249       & (1--0)R16 & 0.674 (a)     & -3.969(a)   & ...      & 5.869 (c) \\
           &                 &           &               &             &          &           \\
Fe I       & 19923.343       & ...       & 5.020         & -1.530      & -7.16    & ...       \\
           & 20349.718       & ...       & 4.186         & -2.625      & -7.54    & ...       \\
           & 20840.808       & ...       & 6.027         &  0.075      & -6.94    & ...       \\
           & 20991.042       & ...       & 4.143         & -3.150      & -7.73    & ...       \\
           & 21238.467       & ...       & 4.956         & -1.300      & -7.00    & ...       \\
           & 21284.344       & ...       & 5.669         & -3.320      & -7.87    & ...       \\
           & 21284.348       & ...       & 3.071         & -4.470      & -7.27    & ...       \\
           & 21735.462       & ...       & 6.175         & -0.700      & -6.87    & ...       \\
           & 22257.107       & ...       & 5.064         & -0.745      & -7.00    & ...       \\
           & 22260.180       & ...       & 5.086         & -0.941      & -7.54    & ...       \\
           & 22380.797       & ...       & 5.033         & -0.600      & -6.99    & ...       \\
           & 22385.102       & ...       & 5.320         & -1.550      & -7.00    & ...       \\
           & 22392.879       & ...       & 5.100         & -1.270      & -7.10    & ...       \\
           & 22419.982       & ...       & 6.218         & -0.234      & -6.85    & ...       \\
           & 22419.990       & ...       & 6.283         & -4.568      & -8.10    & ...       \\
           & 22473.268       & ...       & 6.119         &  0.381      & -6.88    & ...       \\
           & 23308.477       & ...       & 4.076         & -2.645      & -7.18    & ...       \\
           & 23566.666       & ...       & 6.144         &  0.197      & -7.09    & ...       \\
           & 23683.741       & ...       & 5.305         & -1.152      & -7.50    & ...       \\
\hline
\end{tabular}
\begin{tablenotes}
\item \textbf{Notes}: (a) J\"onsson et al. (2014a). (b) Decin (2010). (c) Sauval \& Tatum (1984).
\end{tablenotes}
\label{tab:lines}
\end{table}

\clearpage

\begin{sidewaystable}
	\scalefont{0.65}
	\centering
	\caption{Individual Fe I line abundances from infrared spectra}
	\begin{tabular}{lllllllllllllllllll}
		\hhline{===================}
		Star                              & A(Fe)       & A(Fe)       & A(Fe)       & A(Fe)       & A(Fe)       & A(Fe)       & A(Fe)       & A(Fe)       & A(Fe)       & A(Fe)          & A(Fe)          & A(Fe)          & A(Fe)          & A(Fe)          & A(Fe)          & A(Fe)          & A(Fe) &    \\
		\multicolumn{1}{l}{ }             & $\lambda_1$ & $\lambda_2$ & $\lambda_3$ & $\lambda_4$ & $\lambda_5$ & $\lambda_6$ & $\lambda_7$ & $\lambda_8$ & $\lambda_9$ & $\lambda_{10}$ & $\lambda_{11}$ & $\lambda_{12}$ & $\lambda_{13}$ & $\lambda_{14}$ & $\lambda_{15}$ & $\lambda_{16}$ & $\lambda_{17}$ &  \\ \hline
		Arcturus                          & 6.97        & 6.93        & 6.91        & 7.00        & 6.92        & 6.82        & 6.92        & 6.84        & 6.94        & 6.86           & 7.01           & 6.88           & 6.95           & 6.82           & 7.02           & ...            & ...            &  \\
		10 Dra                            & ...         & ...         & ...         & 7.37        & ...         & ...         & 7.42        & ...         & ...         & 7.37           & ...            & ...            & 7.48           & 7.34           & 7.34           & ...            & ...            &  \\
		41 Com                            & ...         & ...         & ...         & 7.42        & ...         & 7.38        & 7.43        & 7.28        & ...         & 7.32           & ...            & 7.32           & 7.43           & 7.44           & 7.27           & ...            & ...            &  \\
		$\beta$ And                       & 7.48        & ...         & 7.38        & ...         & 7.37        & ...         & 7.44        & ...         & 7.37        & 7.34           & 7.35           & ...            & ...            & 7.34           & 7.39           & ...            & ...            &  \\
		$\beta$ Peg                       & ...         & ...         & ...         & 7.27        & ...         & ...         & ...         & 7.03        & ...         & 7.14           & 7.23           & ...            & 7.18           & 7.28           & 7.17           & ...            & ...            &  \\
		$\beta$ UMi                       & ...         & ...         & ...         & ...         & ...         & ...         & ...         & ...         & ...         & 7.42           & ...            & 7.37           & 7.48           & 7.34           & 7.45           & ...            & ...            &  \\
		$\gamma$ Dra                      & 7.40        & ...         & 7.40        & 7.49        & 7.36        & 7.35        & ...         & 7.33        & 7.35        & 7.32           & ...            & ...            & ...            & 7.34           & 7.43           & ...            & ...            &  \\
		$\mu$ Leo                         & 7.75        & ...         & ...         & ...         & ...         & 7.65        & ...         & 7.58        & ...         & ...            & ...            & ...            & 7.83           & ...            & 7.94           & ...            & ...            & \\
		$\omega$ Vir                      & ...         & 7.46        & ...         & 7.54        & ...         & ...         & 7.45        & 7.35        & ...         & 7.44           & ...            & ...            & 7.50           & 7.48           & 7.34           & ...            & ...            &  \\
		BD$+$06$^\circ$2063               & ...         & ...         & ...         & ...         & ...         & ...         & ...         & ...         & 7.28        & 7.32           & ...            & 7.22           & ...            & 7.39           & 7.49           & ...            & ...            &  \\
		BD$+$16$^\circ$3426               & ...         & ...         & ...         & 7.49        & ...         & ...         & 7.48        & 7.31        & ...         & ...            & ...            & ...            & 7.56           & 7.44           & 7.34           & ...            & ...            &  \\
		HD 96360                          & ...         & 7.31        & ...         & ...         & 7.36        & ...         & ...         & ...         & ...         & 7.27           & ...            & 7.22           & 7.50           & 7.29           & 7.27           & ...            & ...            &  \\
		HD 189581                         & ...         & ...         & ...         & 7.17        & 7.06        & ...         & ...         & 7.13        & 7.15        & ...            & ...            & ...            & ...            & 7.27           & 6.94           & ...            & ...            &  \\
		TYC 5880-423-1                    & ...         & ...         & ...         & ...         & ...         & ...         & ...         & ...         & ...         & ...            & ...            & ...            & ...            & ...            & 6.64           & ...            & ...            & \\
		UCAC2 22625431                    & ...         & ...         & ...         & ...         & ...         & ...         & ...         & ...         & ...         & ...            & ...            & ...            & ...            & ...            & 6.82           & ...            & ...            & \\
		2M06232713$-$3342412              & ...         & ...         & ...         & ...         & ...         & ...         & ...         & ...         & ...         & ...            & ...            & ...            & ...            & ...            & 6.98           & ...            & ...            & \\
		2M07313775$-$2818395              & ...         & ...         & ...         & ...         & ...         & ...         & ...         & ...         & ...         & ...            & ...            & ...            & ...            & ...            & 7.02           & ...            & ...            & \\
		2M01051470$+$4958078              & ...         & ...         & ...         & ...         & ...         & ...         & ...         & ...         & ...         & ...            & ...            & ...            & ...            & ...            & 6.42           & 6.39           & 6.47           &  \\
		2M02431985$+$5227501              & ...         & ...         & ...         & ...         & ...         & ...         & ...         & ...         & ...         & ...            & ...            & ...            & ...            & ...            & 6.59           & ...            & ...            & \\
		2M04224371$+$1729196              & ...         & ...         & ...         & ...         & ...         & ...         & ...         & ...         & ...         & ...            & ...            & ...            & ...            & ...            & 6.61           & 6.67           & ...            &  \\
		2M20410375$+$0001223              & ...         & ...         & ...         & ...         & ...         & ...         & ...         & ...         & ...         & ...            & ...            & ...            & ...            & ...            & 6.56           & ...            & 6.66           & \\
		2M22341156$+$4425220              & ...         & ...         & ...         & ...         & ...         & ...         & ...         & ...         & ...         & ...            & ...            & ...            & ...            & ...            & ...            & 6.23           & 6.19           &  \\ \hline
	\end{tabular}
	\begin{tablenotes}
		\item \textbf{Notes}: Iron lines: $\lambda_1$ = 19923 \AA, $\lambda_2$ = 20349 \AA, $\lambda_3$ = 20840 \AA, $\lambda_4$ = 20991 \AA, $\lambda_5$ = 21238 \AA, $\lambda_6$ = 21284 \AA, $\lambda_7$ = 21735 \AA, $\lambda_8$ = 22257 \AA, $\lambda_9$ = 22260 \AA, $\lambda_{10}$ = 22380 \AA, $\lambda_{11}$ = 22385 \AA, $\lambda_{12}$ = 22392 \AA, $\lambda_{13}$ = 22419 \AA, $\lambda_{14}$ = 22473 \AA, $\lambda_{15}$ = 23308 \AA, $\lambda_{16}$ = 23566 \AA and $\lambda_{17}$ = 23683 \AA.
	\end{tablenotes}
	\label{tab:abundance}
\end{sidewaystable}

\clearpage

\clearpage

\begin{table}
\scalefont{0.9}
    \centering
    \caption{Individual HF line Fluorine abundances}
   \begin{tabular}{lllllll}
   \hhline{=======}
Star                   & A(F)$_{R9}$ & A(F)$_{R13}$ & A(F)$_{R14}$ & A(F)$_{R15}$ & A(F)$_{R16}$ &  \\ \hline
Arcturus               & 3.77        & 3.50         & ...          & ...          & ...          &  \\
10 Dra                 & 4.16        & 4.00         & 4.18         & ...          & 4.05         & \\
41 Com                 & 4.24        & 4.14         & ...          & 4.32         & 4.18         &  \\
$\beta$ And            & 4.40        & ...          & 4.40         & ...          & ...          &  \\
$\beta$ Peg            & 4.05        & ...          & 4.10         & ...          & 3.97         &  \\
$\beta$ UMi            & 4.25        & 4.15         & ...          & ...          & ...          &  \\
$\gamma$ Dra           & 4.45        & ...          & ...          & ...          & ...          &         \\
$\omega$ Vir           & 4.48       & 4.25          & 4.48         & 4.35         & 4.27         &  \\
BD$+$06$^\circ$2063    & 4.30       & 4.33          & ...          & ...          & 4.33         &  \\
BD$+$16$^\circ$3426    & 4.55       & ...           & 4.45         & 4.65         & 4.45         &  \\
HD 96360               & 4.46       & 4.46          & 4.42         & ...          & ...          &  \\
HD 189581              & 4.35       & ...           & 4.25         & ...          & ...          &  \\
HD 19697               & 3.55       & ...           & ...          & ...          & ...          &          \\
HD 20305               & 3.58       & ...           & ...          & ...          & ...          &           \\
HD 90862               & 3.47       & ...           & ...          & ...          & ...          &          \\
TYC 5880-423-1         & 3.35       & ...           & ...          & ...          & ...          &           \\
UCAC2 22625431         & 3.49       & ...           & ...          & ...          & ...          &           \\
2M J06232713$-$3342412 & 3.62       & ...           & ...          & ...          & ...          &           \\
2M J07313775$-$2818395 & 4.30       & ...           & ...          & ...          & ...          &           \\
2M J01051470$+$4958078 & 3.06       & ...           & ...          & ...          & ...          &           \\
2M J02431985$+$5227501 & 3.75       & ...           & ...          & ...          & ...          &           \\
2M J04224371$+$1729196 & 4.14       & ...           & ...          & ...          & ...          &           \\
2M J20410375$+$0001223 & 3.10       & ...           & ...          & ...          & ...          &           \\
2M J22341156$+$4425220 & 2.87       & ...           & ...          & ...          & ...          &           \\
\hline
\end{tabular}
\begin{tablenotes}
\item \textbf{Notes}: HD 28085 not included due to absence of HF lines. \end{tablenotes}
\label{tab:abundance2}
\end{table}

\clearpage

\begin{table}
\scalefont{0.65}
\centering
\caption{Optical Iron lines and Equivalent Widths}
\begin{tabular}{cccccccccccc}
\hhline{============}
$\lambda_{air}$ & $\chi$ &          &            & \multicolumn{2}{c}{HD 19697} & \multicolumn{2}{c}{HD 20305} & \multicolumn{2}{c}{HD 28085} & \multicolumn{2}{c}{HD 90862} \\ 
(\AA)           & (eV)   & log $gf$ & $\Gamma_6$ & EW (m\AA)  & A(Fe)           & EW (m\AA)  & A(Fe)           & EW (m\AA)  & A(Fe)           & EW (m\AA)  & A(Fe)           \\ \hline
6056.005        & 4.73   & -0.556   & -7.130     & 78.2       & 7.07            & 92.2        & 7.11           & 102.       & 7.45            & 85.3       & 6.92           \\
6078.491        & 4.80   & -0.292   & -7.410     & 77.0       & 6.88            & 94.6        & 7.00           & 106.       & 7.37            & 91.2       & 6.87           \\
6079.009        & 4.65   & -1.064   & -7.177     & 52.6       & 6.98            & 68.7        & 7.10           & 73.6       & 7.37           & 50.7       & 6.71            \\
6082.710        & 2.22   & -3.622   & -7.654     & 118.       & 7.36            & 117.        & 7.18           & 96.6       & 7.36           & 127.       & 7.20            \\
6093.642        & 4.61   & -1.402   & -7.202     & 41.2       & 7.02            & 45.8        & 6.97           & 58.3       & 7.40           & 41.1       & 6.81            \\
6096.664        & 3.98   & -1.861   & -7.152     & 60.8       & 6.98            & 61.4        & 6.85           & 70.1       & 7.32           & 61.6       & 6.75            \\
6098.243        & 4.56   & -1.825   & -7.238     & 39.5       & 7.34            & 38.3        & 7.19           & 38.7       & 7.44           & 37.3       & 7.13            \\
6127.906        & 4.14   & -1.503   & -7.790     & 74.5       & 7.09            & 83.9        & 7.08           & 91.4       & 7.49           & 66.9       & 6.70            \\
6136.993        & 2.20   & -3.037   & -7.691     & 144.       & 7.19            & 150.        & 7.12           & 143.       & 7.65           & 162.       & 7.06            \\
6151.618        & 2.18   & -3.357   & -7.696     & 129.       & 7.19            & 130.        & 7.06           & 111.       & 7.27           & 137.       & 6.95            \\
6157.728        & 4.08   & -1.257   & -7.790     & 109.       & 7.45            & 32.7        & 7.23           & 120.       & 7.75           & 114.       & 7.23            \\
6159.378        & 4.61   & -1.910   & -7.216     & ...        & ...             & ...         & ...            & 34.3       & 7.50           & 44.8       & 7.39            \\
6165.360        & 4.14   & -1.487   & -7.780     & 69.1       & 6.97            & 77.1        & 6.95           & 80.1       & 7.28           & 75.8       & 6.83            \\
6173.336        & 2.22   & -2.938   & -7.690     & 147.       & 7.16            & 152.        & 7.07           & 137.       & 7.48           & 158.       & 6.92            \\
6187.990        & 3.94   & -1.724   & -7.179     & 86.6       & 7.26            & 83.7        & 7.03           & 82.2       & 7.32           & 81.1       & 6.88            \\
6200.312        & 2.61   & -2.457   & -7.589     & 121.       & 6.78            & 140.        & 6.99           & 134.       & 7.45           & 142.       & 6.82            \\
6213.429        & 2.22   & -2.650   & -7.691     & 160.       & 7.05            & 171.        & 7.01           & 157.       & 7.50           & 180.       & 6.87            \\
6219.280        & 2.20   & -2.549   & -7.694     & 178.       & 7.11            & 187.        & 7.03           & 166.       & 7.49           & 199.       & 6.93            \\
6220.779        & 3.88   & -2.390   & -7.208     & 63.7       & 7.41            & 61.5        & 7.24           & 45.8       & 7.34           & 33.7       & 6.64            \\
6226.734        & 3.88   & -2.143   & -7.208     & 76.4       & 7.40            & 62.8        & 7.02           & 66.5       & 7.42           & 58.3       & 6.83            \\
6232.640        & 3.65   & -1.232   & -7.540     & 112.       & 6.88            & 126.        & 6.94           & 129.       & 7.38           & 122.       & 6.75            \\
6240.646        & 2.22   & -3.337   & -7.661     & 132.       & 7.52            & 115.        & 6.83           & 108.       & 7.29           & 140.       & 7.14            \\
6265.132        & 2.18   & -2.633   & -7.700     & 190.       & 7.25            & 186.        & 7.06           & 171.       & 7.58           & 202.       & 6.99            \\
6335.329        & 2.20   & -2.423   & -7.698     & 189.       & 7.04            & ...         & ...            & 165.       & 7.30           & 199.       & 6.77            \\
6380.743        & 4.19   & -1.312   & -7.790     & 84.0       & 7.14            & 86.6        & 7.00           & 89.7       & 7.31           & 90.2       & 6.98            \\
6385.718        & 4.73   & -1.887   & -7.187     & 18.8       & 7.11            & 20.6        & 7.08           & 29.2       & 7.51           & 19.7       & 6.99            \\
6392.537        & 2.28   & -4.007   & -7.643     & 79.2       & 6.91            & 78.5        & 6.86           & 62.2       & 7.27           & 84.2       & 6.74            \\
6574.226        & 0.99   & -5.019   & -7.830     & 157.       & 7.74            & 144.        & 7.20           & 104.       & 7.26           & 164.       & 7.20            \\
6591.312        & 4.59   & -2.065   & -7.697     & 35.7       & 7.54            & 25.6        & 7.20           & 28.4       & 7.50           & 26.1       & 7.13            \\
6593.871        & 2.43   & -2.469   & -7.629     & 154.       & 7.13            & 174.        & 7.11           & ...        & ...            & 174.       & 6.91            \\
6597.561        & 4.80   & -0.984   & -7.190     & 57.1       & 7.18            & 54.3        & 6.95           & 72.2       & 7.41           & 57.9       & 6.96            \\
6608.025        & 2.28   & -4.017   & -7.648     & 85.4       & 7.01            & 88.0        & 7.00           & 66.9       & 7.33           & 91.6       & 6.85            \\
6609.110        & 2.56   & -2.708   & -7.610     & 136.       & 7.28            & 143.        & 7.13           & 144.       & 7.71           & 149.       & 6.99            \\
6627.543        & 4.55   & -1.542   & -7.250     & 43.5       & 7.11            & 50.2        & 7.10           & 61.6       & 7.49           & 54.3       & 7.09            \\
6646.930        & 2.61   & -3.988   & -7.604     & 66.5       & 7.16            & 61.9        & 7.05           & 55.5       & 7.54           & 71.2       & 7.00            \\
6653.851        & 4.15   & -2.475   & -7.153     & 32.7       & 7.25            & 23.9        & 6.98           & 28.2       & 7.39           & 30.6       & 7.03            \\
6699.141        & 4.59   & -2.167   & -7.667     & 31.8       & 7.55            & 22.3        & 7.22           & 23.3       & 7.48           & 20.8       & 7.10            \\
6703.565        & 2.76   & -3.078   & -7.633     & 101.       & 7.14            & 105.        & 7.08           & 94.1       & 7.38           & 109.       & 6.97            \\
6710.318        & 1.49   & -4.876   & -7.733     & 108.       & 7.12            & 106.        & 7.01           & 87.6       & 7.49           & 120.       & 7.04            \\
6713.742        & 4.80   & -1.485   & -7.207     & 38.4       & 7.29            & 29.0        & 6.96           & 41.6       & 7.39           & 34.3       & 7.01            \\
6725.355        & 4.10   & -2.257   & -7.181     & 28.0       & 6.85            & 40.1        & 7.02           & 42.9       & 7.38           & 35.4       & 6.84            \\
6726.666        & 4.61   & -1.062   & -7.500     & 56.9       & 6.98            & 67.6        & 7.00           & 71.9       & 7.24           & 58.5       & 6.77            \\
6732.064        & 4.58   & -2.208   & -7.700     & 40.5       & 7.77            & ...         & ...            & 25.0       & 7.55           & 25.9       & 7.26            \\
6733.151        & 4.64   & -1.490   & -7.247     & 49.4       & 7.31            & 44.6        & 7.07           & 58.6       & 7.50           & 47.6       & 7.05            \\
6739.520        & 1.56   & -4.955   & -7.726     & 97.5       & 7.08            & 85.0        & 6.84           & 57.5       & 7.24           & 90.5       & 6.68            \\
6745.100        & 4.58   & -2.164   & -7.726     & 29.5       & 7.47            & 16.9        & 7.04           & 19.7       & 7.37           & 17.7       & 6.99            \\
6745.955        & 4.08   & -2.709   & -7.820     & 19.8       & 7.05            & 15.1        & 6.85           & 18.0       & 7.28           & 13.4       & 6.69            \\
6750.151        & 2.42   & -2.672   & -7.609     & 153.       & 7.42            & 158.        & 7.13           & 154.       & 7.67           & 165.       & 7.00            \\
6806.842        & 2.73   & -3.153   & -7.643     & 106.       & 7.28            & 101.        & 7.02           & 95.4       & 7.35           & 111.       & 7.02            \\
6839.830        & 2.60   & -3.433   & -7.617     & 95.5       & 7.11            & 102.        & 7.13           & 86.6       & 7.41           & 106.       & 7.00            \\
6842.685        & 4.64   & -1.245   & -7.189     & 58.8       & 7.24            & 65.7        & 7.18           & 67.9       & 7.39           & 52.6       & 6.89            \\
6857.249        & 4.08   & -2.125   & -7.820     & 49.1       & 7.13            & 46.9        & 6.97           & ...        & ...            & 50.0       & 6.93            \\
\hline
\end{tabular}
\label{tab:lines2}
\end{table}

\clearpage

\begin{table}
\centering
\caption{Abundance Sensitivities to Stellar Parameters}
\begin{tabular}{ccccccl}
\hhline{=======}
Species         & $\delta$T$\rm_{eff}$=+100 K & $\delta\log$g=+0.25 & $\delta$[Fe/H]=+0.1 & $\delta\xi$=+0.3 km$\cdot$s$^{-1}$ & $\Delta$A  \\ \hline
HF              & +0.15 & -0.01 & +0.03 & -0.04 & $\pm$ 0.16 $^{*}$  \\
HF              & +0.18 & +0.00 & +0.08 & +0.00 & $\pm$ 0.18$^{**}$  \\
Fe I (infrared) & -0.01 & +0.04 & +0.00 & -0.04 & $\pm$ 0.06$^{*}$   \\
Fe I (optical)  & -0.01 & +0.06 & +0.00 & -0.12 & $\pm$ 0.13$^{***}$ \\
\hline
\end{tabular}
\begin{tablenotes}
\item \textbf{Notes}: * Baseline model: T$\rm _{eff}$ = 3917 K; log g = 1.30; $\xi$ = 1.75 km/s;  [Fe/H]=-0.08.

** Baseline model: T$\rm _{eff}$ = 3725 K; log g = 0.30; $\xi$ = 1.70 km/s; [Fe/H] = -1.10;

*** Baseline model: T$\rm _{eff}$ = 4075 K; log g = 1.45; $\xi$ = 1.67 km/s; [Fe/H] = -0.40;
\end{tablenotes}
\label{tab:disturbance}
\end{table}

\clearpage

\clearpage

\clearpage

\clearpage

\clearpage

\clearpage

\tablewidth{0pt}	


\begin{thebibliography}{}

\bibitem[Abia et al. (2009)]{abia2009} Abia, C., Recio-Blanco, A., de Laverny, P., et al. 2009, ApJ, 694, 971 

\bibitem[Abia et al. (2010)]{abia2010} Abia, C., Cunha, K., Cristallo, S., et al. 2010, ApJL, 715, L94 

\bibitem[Abia et al. (2015)]{abia2015} Abia, C., Cunha, K., Cristallo, S. \& de Laverny, P. 2015, A\&A 581, A88 

\bibitem[Abia et al. (2019)]{abia2019} Abia1, C., Cristallo, S., Cunha, K., de Laverny, P. \& Smith, V. V. 2019, A\&A 625, A40 

\bibitem[Alib\'es et al. (2001)]{alibes2001} Alib\'es, A., Labay, J., Canal, R. 2001, A\&A, 370, 1103 

\bibitem[Anders et al. (2017)]{anders2017} Anders, F., Chiappini, C., Minchev, I., Miglio, A., et al. 2017, A\&A 600, A70 

\bibitem[Asplund et al. (2005)]{asplund2005} Asplund, M., Grevesse, N., \& Sauval, A. J. 2005, in ASP Conf. Ser. 336, Cosmic Abundances as Records of Stellar Evolution and Nucleosynthesis, ed. T. G. Barnes III \& F. N. Bash (San Francisco: ASP), 25 

\bibitem[Asplund et al.(2009)]{Asplund2009} Asplund, M., Grevesse, N., Sauval, A. J., \& Scott, P. 2009, ARA\&A, 47, 481 

\bibitem[Alvarez \& Plez (1998)]{Alvarez_Plez1998}  Alvarez, R., \& Plez, B. 1998, A\&A, 330, 1109 

\bibitem[Alves-Brito et al. (2010)]{alvesbrito2010} Alves-Brito, A., Mel\'endez, J., Asplund, M., Ram\'irez, I., \& Yong, D. 2010, A\&A, 513, A35 

\bibitem[Alves-Brito et al. (2012)]{alvesbrito2012} Alves-Brito, A., Yong, D., Mel\'endez, J., V\'asquez, S., \& Karakas, A. I. 2012, A\&A, 540, A3 

\bibitem[Auri\'ere (2003)]{Auriere2003} Auri\'ere, M. 2003, EAS Publications Series, 9, 105 

\bibitem[Bailer-Jones et al. (2018)]{Bailer-Jones2018} Bailer-Jones, C. A. L., Rybizki, J., Fouesneau, M., Mantelet, G., \& Andrae, R. 2018, AJ, 156, 58 

\bibitem[Belokurov et al.(2018)]{belokurov2018} Belokurov, V., Erkal, D., Evans, N.~W., et al.\ 2018, \mnras, 478, 611 

\bibitem[Bergemann et al. (2014)]{bergemann2014} Bergemann, M., Ruchti, G. R., Serenelli, A., Feltzing, S., et al. 2014, A\&A 565, A89 

\bibitem[Bessell et al. (1998)]{bessell1998} Bessell, M. S., Castelli, F., \& Plez, B. 1998, A\&A, 333, 231 

\bibitem[Bizyaev et al. (2006)]{bizyaev2006} Bizyaev, D., Smith, V. V., Arenas, J., et al. 2006, AJ, 131, 1784 

\bibitem[Boeche et al. (2013)]{boeche2013} Boeche, C., Siebert, A.. Piffl, T., Just, A., et al. 2013, A\&A, 559, A59 

\bibitem[Bovy et al. (2016)]{bovy2016} Bovy, J., Rix, H.-W., Schlafly, E. F., et al. 2016, ApJ, 823, 30 

\bibitem[Bragan\c{c}a et al. (2019)]{graganca2019} Bragan\c{c}a, G. A., Daflon, S., Lanz, T., Cunha, K., et al. 2019, A\&A 625, A120 

\bibitem[Bressan et al. (2012)]{bressan2012} Bressan, A., Marigo, P., Girardi, L., et al. 2012, \mnras, 427, 127 

\bibitem[Carpenter (2001)]{carpenter2001} Carpenter, J. M. 2001, AJ, 121, 2851 

\bibitem[Carrera et al. (2019)]{carrera2019} Carrera, R., Bragaglia, A., Cantat-Gaudin, T., Vallenari, A., et al. 2019, A\&A, 623, A80 

\bibitem[Cayrel (1988)]{cayrel1988} Cayrel, R. 1988, in IAU Symp. 132, The Impact of Very High S/N Spectroscopy on Stellar Physics, ed. R. Cayrel, G. de Strobel, \& M. Spite ( Dordrecht: Kluwer), 354

\bibitem[Chen et al. (1998a)]{Chen1998a} Chen, B., Vergely, J. L., Valette, B., \& Carraro, G. 1998a, A\&A, 336, 137 

\bibitem[Chen et al. (1998b)]{Chen1998b} Chen, P. S., Wang, X. H., \& Xiong, G. Z. 1998b, A\&A, 333, 613 

\bibitem[Cheng et al. (2012)]{cheng2012} Cheng, J. Y., Rockosi, C. M., Morrison, H. L., Sch\"onrich, R. A., et al. 2012, \apj, 746, 149 

\bibitem[Chou et al.(2010)]{chou2010} Chou, M.-Y., Majewski, S.~R., Cunha, K., et al.\ 2010, \apj, 720, L5 

\bibitem[Crane et al.(2003)]{crane2003} Crane, J.~D., Majewski, S.~R., Rocha-Pinto, H.~J., et al.\ 2003, \apj, 594, L119 

\bibitem[Cristallo et al. (2014)]{cristallo2014} Cristallo, S., Di Leva, A., Imbriani, G., et al. 2014, A\&A, 570, A46 

\bibitem[Cunha et al. (2003)]{cunha2003} Cunha, K., Smith, V. V., Lambert, D. L. \& Hinkle, K. H 2003, AJ, 126, 1305 

\bibitem[Cushing, Vacca \& Rayner (2004)]{cushing2004} Cushing, M. C., Vacca, W. D. \& Rayner, J. T. 2004, PASP, 116, 362 

\bibitem[da Silva et al. (2006)]{dasilva2006} da Silva, L., Girardi, L., Pasquini, L., Setiawan, J., von der L\"uhe, O., de Medeiros, J. R., Hatzes, A., D\"ollinger, M. P. \& Weiss, A. 2006, A\&A 458, 609 

\bibitem[Daflon \& Cunha (2004)]{daflon_cunha2004} Daflon, S., \& Cunha, K. 2004, \apj, 617, 1115 

\bibitem[de Laverny \& Recio-Blanco (2013a)]{delaverny_recioblaco2013a} de Laverny, P. \& Recio-Blanco, A. 2013a, A\&A, 555, A121 

\bibitem[de Laverny \& Recio-Blanco (2013b)]{delaverny_recioblaco2013b} de Laverny, P., \& Recio-Blanco, A. 2013b, A\&A, 560, A74 

\bibitem[Decin (2000)]{decin2000} Decin, L. 2000, PhD thesis, Catholique Univ. Leuven 16 

\bibitem[Donor et al. (2018)]{donor2018} Donor, J., Frinchaboy, P. M., Cunha, K., Thompson, B., et al. 2018, AJ, 156, 142 

\bibitem[D'Orazi et al. (2013)]{dorazi2013} D'Orazi, V., Lucatello, S., Lugaro, M. Gratton, R. G., Angelou, G., Bragaglia, A., Carretta, E., Alves-Brito, A. \& Ivan, I. I. 2013, ApJ, 763, 22 

\bibitem[Esteban \& Garc\'ia-Rojas (2018)]{estebangarcia2018} Esteban, C., \& Garc\'ia-Rojas, J. 2018, MNRAS, 478, 2315 

\bibitem[Fenner et al. (2004)]{fenner2004} Fenner, Y., Campbell, S., Karakas, A. I., Lattanzio, J. C., \& Gibson, B. K. 2004, MNRAS, 353, 789

\bibitem[Foresti et al. (1992)]{foresti1992} Forestini, M., Goriely, S., Jorissen, A., \& Arnould, M. 1992, A\&A, 261, 157 

\bibitem[Gaia Collaboration (2018)]{GaiaCollaboration2018} Gaia Collaboration, Brown, A. G. A., Vallenari, A., et al. 2018, A\&A 616, A1 

\bibitem[Garc\'ia P\'erez et al. (1016)]{garcia_perez2016} Garc\'ia P\'erez, A. E., Prieto, C. A., Holtzman, J. A., Shetrone, M., et al. 2016, AJ, 151, 144 

\bibitem[Genovali et al. (2014)]{genovali2014} Genovali, K., Lemasle, B., Bono, G., Romaniello, M., et al. 2014, A\&A, 566, A37 

\bibitem[Ghezzi et al. (2018)]{ghezzi2018} Ghezzi, L., Montet, B. T., \& Johnson, J. A. 2018, ApJ, 860, 109 

\bibitem[Gillessen et al. (2009)]{gillessen2009} Gillessen, S., Eisenhauer, F., Trippe, S., et al. 2009, ApJ, 692, 1075 

\bibitem[Goriely et al. (19990)]{goriely1990} Goriely S., Jorissen A., Arnould M., 1990, in Hillebrandt W., M\"uller E., eds, Nuclear Astrophysics 5th Workshop. Max Planck Institut f\"ur As- trophysik, Ringberg Castle, Germany, p. 60 

\bibitem[Green et al. (2018)]{green2018} Green, G. M., Schlafly, E. F., Finkbeiner, D., et al. 2018, \mnras, 478, 651 

\bibitem[Guer\c{c}o et al. (2019)]{guerco2019} Guer\c{c}o, R., Cunha, K., Smith, V. V., Pereira, C. B., Abia, C., Lambert, D. L., de Laverny, P., Recio-Blanco, A. \& J\"onsson, H. 2019, ApJ, 876, 43 

\bibitem[Gustafsson et al. (2008)]{gustafsson2008} Gustafsson B., Edvardsson B., Eriksson K., et al. 2008, A\&A 486, 951 

\bibitem[Hall el al. (1979)]{hall1979} Hall, D. N. B., Ridgway, S. T., Bell, E. A., \& Yarborough, J. M. 1979, Proc. SPIE, 172, 121

\bibitem[Hayes et al.(2018)]{hayes2018} Hayes, C.~R., Majewski, S.~R., Shetrone, M., et al.\ 2018, \apj, 852, 49 

\bibitem[Hayden et al. (2015)]{hayden2015} Hayden, M. R., Bovy, J., Holtzman, J. A., Nidever, D. L., et al. 2015, ApJ, 808, 132 

\bibitem[Heger et al. (2005)]{heger2005} Heger, A., Kolbe, E., Haxton, W. C., Langanke, K., Martinez-Pinedo, G., Woosley, S. E. 2005, PhLB, 606, 258

\bibitem[Helmi et al.(2018)]{helmi2018} Helmi, A., Babusiaux, C., Koppelman, H.~H., et al.\ 2018, \nat, 563, 85 

\bibitem[Hinkle et al. (2000)]{hinkle2000} Hinkle, K., Wallace, L., Valenti, J. \& Harmer, D. 2000, Astronomical Society of the Pacific Monograph Publication, ``Visible and Near Infrared Atlas of the Arcturus Spectrum'', vol. 2 (\url{http://aspmonographs.org/a/volumes/table_of_contents/?book_id=5}) 

\bibitem[Hinkle et al. (2003)]{hinkle2003} Hinkle, K. H., Blum, R. D., Joyce, R. R., Sharp, N., Ridgway, S. T., van der Bliek, N. S., Rogers, B., Smith, V. \& Valenti, J. 2003, SPIE, 4834, 353

\bibitem[Hinkle \& Lambert (1975)]{} Hinkle, K. H. \& Lambert, D. L. 1975, MNRAS, 170, 447 

\bibitem[Hinkle, Wallace & Livingston (1995)]{hinkle1995} Hinkle, K., Wallace, L., \& Livingston, W. 1995, PASP, 107, 1042 

\bibitem[Holtzman et al.(2018)]{holtzman2018} Holtzman, J.~A., Hasselquist, S., Shetrone, M., et al.\ 2018, \aj, 156, 125 

\bibitem[Ibata et al.(2003)]{ibata2003} Ibata, R.~A., Irwin, M.~J., Lewis, G.~F., et al.\ 2003, \mnras, 340, L21 

\bibitem[J\"onsson et al. (2014a)]{jonsson2014a} J\"onsson, H., Ryde, N., Harper, G. M., Cunha, K., Schultheis, M., Eriksson, K., Kobayashi, C., Smith, V. V., \& Zoccali, M. 2014a, A\&A, 564, A122 

\bibitem[J\"onsson et al. (2014b)]{jonsson2014b} J\"onsson, H., Ryde, N., Harper, G. M., Richter, M. J., \& Hinkle, K. H. 2014b, ApJL, 789, L41 

\bibitem[J\"onsson et al. (2017)]{jonsson2017} J\"onsson, H., Ryde, N., Spitoni, E., et al. 2017, ApJ, 835, 50 

\bibitem[Jorissen et al. (1992)]{jorissen1992} Jorissen, A., Smith, V. V. \& Lambert, D. L. 1992, A\&A, 261, 164 

\bibitem[Kobayashi et al. (2006)]{kobayashi2006} Kobayashi, C., Umeda, H., Nomoto, K., Tominaga, N., \& Ohkubo, T. 2006, ApJ, 653, 1145 

\bibitem[Kobayashi et al. (2011a)]{kobayashi2011a} Kobayashi, C., Karakas, A. I. \& Umeda, H. 2011, MNRAS, 414, 3231 

\bibitem[Kobayashi et al. (2011b)]{kobayashi2011b} Kobayashi1, C., Izutani, N., Karakas, A. I., Yoshida, T. Yong, D. \& Umeda, H. 2011, ApJ, 739, L57 

\bibitem[Kordopatis et al. (2013)]{kordopatis2013} Kordopatis, G., Gilmore, G. \& Steinmetz, M. 2013, AJ, 146, 134 

\bibitem[Kubryk et al. (2015)]{kubryk2015} Kubryk, M. Prantzos, N., \& Athanassoula, E. 2015, A\&A 580, A127 

\bibitem[Kundu et al. (2002)]]{kundo2002} Kundu, A., Majewski, S. R., Rhee, J., et al. 2002, ApJ, 576, L125 

\bibitem[Kurucz (1994)]{kurucz1994} Kurucz, R. L. 1994, Atomic Data for Fe and Ni, Kurucz CD-ROM No. 22. Cambridge, Mass.: Smithsonian Astrophysical Observatory 

\bibitem[Kurucz (2014)]{kurucz2014} Kurucz, R. L. 2014, Robert L. Kurucz on-line database of observed and predicted atomic transitions 

\bibitem[Lemasle et al. (2008)]{lemasle2008} Lemasle, B., François, P., Piersimoni, A., Pedicelli, S, et al. 2008, A\&A, 490, 613 

\bibitem[Li et al. (2013)]{li2013} Li, H. N., Ludwig, H, -G, Caffau, E., Christlieb, N., \& Zhao, G. 2013, 765, 51 

\bibitem[Luo (2007)]{Luo2007} Luo, Y. R. 2007, from ``Comprehensive Handbook of Chemical Bond Energies'' (CRC Press: Boca Raton, Florida  USA) 

\bibitem[Mackereth et al. (2017)]{mackereth2017} Mackereth, J. T., Bovy, J., Schiavon, R. P., et al. 2017, MNRAS, 471, 3057 

\bibitem[Magrini et al. (2017)]{Magrini2017} Magrini, L., Randich, S., Kordopatis, G., Prantzos, N., et al. 2017, A\&A, 603, A2 

\bibitem[Majewski et al. (2017)]{majewski2017} Majewski, S. R., Schiavon, R. P., Frinchaboy, P. M., et al. 2017, AJ, 154, 94 

\bibitem[McCall (2004)]{McCall2004} McCall, M. L. 2004, AJ, 128, 2144 

\bibitem[Meynet \& Arnould (2000)]{Meynet_Arnould (2000)} Meynet, G. \& Arnould, M. 2000, A\&A, 355, 176 

\bibitem[Mikolaitis et al. (2014)]{mikolaitis2014} Mikolaitis, \v{S}., Hill, V., Recio-Blanco, A., de Laverny, P., et al. 2014, A\&A, 572, A33 

\bibitem[Minchev et al. (2013)]{minchev2013} Minchev, I., Chiappini, C., \& Martig, M. 2013, A\&A 558, A9 

\bibitem[Minchev et al. (2015)]{minchev2015} Minchev, I., Martig, M., Streich, D., et al. 2015, ApJL, 804, L9 

\bibitem[Minchev et al. (2017)]{minchev2017} Minchev, I., Steinmetz, M., Chiappini, C., Martig, M., et al. 2017, ApJ, 834, 27 

\bibitem[Morganson et al.(2016)]{morganson2016} Morganson, E., Conn, B., Rix, H.-W., et al.\ 2016, \apj, 825, 140 

\bibitem[Munari et al. (2014)]{Munari2014} Munari, U., Henden, A., Frigo, A., Zwitter, T., et al. 2014, AJ, 148, 81 

\bibitem[Nault \& Pilachowski (2013) ]{nault_Pilachowski2013} Nault, K. A., \& Pilachowski, C. A. 2013, AJ, 146, 153 

\bibitem[Newberg et al. (2002)]{newberg2002} Newberg, H. J., Yanny, B., \& Rockosi, C. 2002, ApJ, 569, 245 

\bibitem[Norris (1986)]{norris1986} Norris, J. 1986, ApJS, 61, 667

\bibitem[Pilachowski et al. (1996)]{pilachowski1996} Pilachowski, C. A., Sneden, C. \& Kraft, R. P. 1996, AJ, 111, 1689 

\bibitem[Pilachowski et al. (2017)]{pilachowski2017} Pilachowski, C. A., Hinkle, K. H., Young, M. D., et al. 2017, PASP, 129, 972 

\bibitem[Pilachowski \& Pace (2015)]{Pilachowski_Pace2015} Pilachowski, C. A. \& Pace, C. 2015, AJ, 150, 66 

\bibitem[Plez (2012)]{plez2012} Plez, B. 2012, Turbospectrum: Code for spectral synthesis, Astrophysics Source Code Library, record ascl:1205.004 

\bibitem[Prantzos et al (2018)]{prantzos2018} Prantzos, N., Abia, C., Limongi, M., Chieffi, A., \& Cristallo, S. 2018, MNRAS, 476, 3432 

\bibitem[Rayner et al. (2016)]{rayner2016} Rayner, J., Tokunaga, A., Jaffe, D., Bonnet, M., et al. 2016, Proc. SPIE, 9908, 990884 

\bibitem[Recio-Blanco et al. (2012)]{recioblanco2012} Recio-Blanco, A., de Laverny, P., Worley, C., Santos, N. C. Melo, C. \& Israelian, G. 2012, A\&A 538, A117 

\bibitem[Renda et al. (2004)]{renda2004} Renda, A., Fenner, Y, Gibson, B. K., Karakas, A. I. et al. 2004, MNRAS, 354, 575 

\bibitem[Rocha-Pinto et al.(2003)]{rochapinto2003} Rocha-Pinto, H.~J., Majewski, S.~R., Skrutskie, M.~F., et al.\ 2003, \apj, 594, L115 

\bibitem[Rodrigues et al. (2014)]{rodrigues2014} Rodrigues, T. S., Girardi, L., Miglio, A., et al. 2014, \mnras, 445, 2758 

\bibitem[Sauval \& Tatum (1984)]{sayval_tatum1984} Sauval, A. J. \& Tatum, J. B. 1984, ApJS, 56, 193 

\bibitem[Sch\"onrich \& Binney (2009)]{Schonrich_binney2009} Sch\"onrich, R., \& Binney, J. 2009, 396, 203 

\bibitem[Sieverding et al. (2018)]{sieverding2018} Sieverding, A., Martinez-Pinedo, G., Huther, L., Langanke, K., Heger, A. 2018, ApJ, 865, 143

\bibitem[Sieverding et al. (2019)]{sieverding2019} Sieverding, A., Langanke, K., Martinez-Pinedo, G., Bollig, R., Janka, H.-T., Heger, A. 2019, ApJ, 876, 151 

\bibitem[Smith et al. (2005)]{smith2005} Smith, V. V., Cunha, K., Ivans, I. I., et al. 2005, ApJ, 633, 392 

\bibitem[Smith et al. (2013)]{smith2013} Smith, V. V., Cunha, K., Shetrone, M. D., Meszaros, S., Prieto, C. A., Bizyaev, D., P\`erez, A. G., Majewski, S. R., Schiavon, R., Holtzman, J., \& Johnson, J. A. 2013, ApJ, 765, 16 

\bibitem[Smith \& Lambert (1985)]{smith_lambert1985} Smith, V. V. \& Lambert, D. L. 1985, ApJ, 294, 326 

\bibitem[Smith \& Lambert (1986)]{smith_lambert1986} Smith, V. V. \& Lambert, D. L. 1986, ApJ, 311, 843 

\bibitem[Smith \& Lambert (1990)]{smith_lambert1990} Smith, V. V. \& Lambert, D. L. 1990, ApJS, 72, 387 

\bibitem[Sneden (1973)]{sneden1973} Sneden, C. 1973, ApJ, 184, 839 

\bibitem[Sneden et al. (2012)]{sneden2012} Sneden, C., Bean, J., Ivans, I., Lucatello, S. \& Sobeck, J. 2012, Astrophysics Source Code Library, record ascl:1202.009 

\bibitem[Spitoni et al. (2018)]{spitoni2018} Spitoni, E., Matteucci, F., J\"onsson, H., Ryde, N. \& Romano, D. 2018, A\&A 612, A16 

\bibitem[Stanghellini \& Haywood (2018)]{stanghellini_Haywood2018} Stanghellini, L., \& Haywood, M. 2018, ApJ, 862, 45 

\bibitem[Timmes et al. (1995)]{timmes1995} Timmes F. X., Woosley S. E., Weaver T. A., 1995, ApJS, 98, 617 

\bibitem[Thorsbro (2016)]{thorsbro2016} Thorsbro, B. 2016, Master thesis, In Lund Observatory Examensarbeten ASTM31 20161 

\bibitem[Tody (1986)]{tody1986} Tody, D. 1986 Proc. SPIE, 0627, Instrumentation in Astronomy VI 

\bibitem[Tody (1993)]{tody1993} Tody, D. 1993, in ASP Conf. Ser. 52, Astronomical Data Analysis Software and Systems II, ed. R. J. Hanisch, R. J. V. Brissenden, \& J. Barnes (San Francisco, CA: ASP), 173 

\bibitem[Uttenthaler et al. (2008)]{uttenthaler2008} Uttenthaler, S., Aringer, B., Lebzelter, T., et al. 2008, ApJ, 682, 509 

\bibitem[Valenti \& Piskunov (1996)]{Valenti_Piskunov1996} Valenti, J. A. \& Piskunov, N. 1996,A\&AS, 118, 595 

\bibitem[van Hoof (2018)]{van_hoof2018} van Hoof, P. A. M. 2018, Galaxies, vol. 6, issue 2, p. 63 

\bibitem[van Leeuwen (2007)]{vanleeuwen2007} van Leeuwen, F. 2007, A\&A, 474, 653 

\bibitem[Weinberg et al. (2019)]{weinberg2019} Weinberg, D. H., Holtzman, J. A., Hasselquist, S., et al. 2019, ApJ, 874, 102 

\bibitem[Woosley \& Haxton (1988)]{Woosley_Haxton1988} Woosley, S. E., \& Haxton, W. C. 1988, Natur, 334, 45 

\bibitem[Woosley et al. (1990)]{woosley1990} Woosley, S. E., Hartmann, D.H., Hoffman, R. D. \& Haxton, W.C. 1990, 356, 272 

\bibitem[Yong et al. (2008)]{yong2008} Yong, D., Mel\'endez, J., Cunha, K., et al. 2008, ApJ, 689, 1020

\bibitem[Yong et al. (2012)]{yong2012} Yong, D., Carney, B. W., \& Friel, E. D. 2012, \aj, 144, 95 

\bibitem[Zacharias et al. (2012)]{Zacharias2012} Zacharias, N., Finch, C. T., Girard, T. M., et al. 2012, VizieR On-line Data Catalog: I/322A. Originally published in: 2012yCat.1322....0Z; 2013AJ....145...44Z 

\end{thebibliography}
\end{document}